\documentclass[final]{WileyASNA-v1}
\usepackage{tensor}
\usepackage{amsmath}
\usepackage{verbatim}
\usepackage{mathbbol}
\usepackage{units}
\usepackage[makeroom]{cancel}
\let\txtstil\relax

\newcommand{\onehalf}{{\txtstil\frac{1}{2}}}
\newcommand{\oneeighth}{{\txtstil\frac{1}{8}}}
\newcommand{\ihalf}{{\txtstil\frac{\rmi}{2}}}
\newcommand{\iquarter}{{\txtstil\frac{\rmi}{4}}}
\newcommand{\dd}{\mathrm{d}}

\newcommand{\pfrac}[2]{\frac{\partial{#1}}{\partial{#2}}}
\newcommand{\ppfrac}[3]{\frac{\partial^{2}{#1}}{\partial{#2}\partial{#3}}}
\newcommand{\detpartial}[2]{\left|\pfrac{#1}{#2}\right|}
\renewcommand{\dfrac}[2]{\frac{\dd#1}{\dd#2}}
\newcommand{\HCd}{\mathcal{H}}
\newcommand{\LCd}{\mathcal{L}}
\newcommand{\FCd}{\mathcal{F}}
\newcommand{\kappabar}{\bar{\kappa}}
\newcommand{\rmi}{\mathrm{i}}
\newcommand{\psibar}{\bar{\psi}}
\newcommand{\Psibar}{\bar{\Psi}}

\newcommand{\KCd}{\mathcal{K}}
\newcommand{\KCdbar}{\bar{\KCd}}
\newcommand{\HO}{\Omega}
\newcommand{\ho}{\omega}
\newcommand{\HOc}{\tilde{Q}}
\newcommand{\hoc}{\tilde{q}}
\newcommand{\must}{\stackrel{!}{=}}
\articletype{Article Type}%
\let\dete\varepsilon\relax
\received{4 March 2021}
\revised {15 April 2021}
\accepted{22 April 2021}
\jyear{2021}
\artid{13991}
\doi{10.1002/asna.202113991}
\relax

\begin{document}

\title{Covariant canonical gauge theory of gravitation for fermions}

\author[1,2]{J\"urgen Struckmeier}
\author[1]{David Vasak}

\authormark{STRUCKMEIER AND VASAK}

\address[1]{\orgname{Frankfurt Institute for Advanced Studies (FIAS)}, \orgaddress{Ruth-Moufang-Str.~1, 60438~Frankfurt am Main}, \country{Germany}}
\address[2]{\orgname{Goethe Universit\"at}, \orgaddress{Max-von-Laue-Str.~1, 60438~Frankfurt~am~Main}, \country{Germany}}
\corres{\email{struckmeier@fias.uni-frankfurt.de (J.S.) and
vasak@fias.uni-frankfurt.de (D.V.)}}

\abstract{%
We derive the interaction of fermions with a dynamical space-time based on the postulate that the description
of physics should be independent of the reference frame,
which means to require the form-invariance of the fermion action under diffeomorphisms.
The derivation is worked out in the Hamiltonian formalism as a canonical transformation along the line of non-Abelian gauge theories.
This yields a closed set of field equations for fermions, unambiguously fixing their coupling to dynamical space-time.
We encounter, in addition to the well-known minimal coupling, anomalous couplings to curvature and torsion.
In torsion-free geometries that anomalous interaction reduces to a Pauli-type coupling with the curvature scalar via
a spontaneously emerged new coupling constant with the dimension of mass resp.\ inverse length.
A consistent model Hamiltonian for the free gravitational field and the impact of its functional form on the structure of the dynamical geometry space-time is discussed.
}


\maketitle

\section{Introduction\label{sec:Intro}}
The early investigations of gauge theories of classical (c-number) fields describing space-time and matter
have been carried out in the Lagrangian
picture~\citep{weyl19,Einstein55,YM54,sciama62,Kibble61,Utiyama56,hehl76}.
In contrast, our approach is based on the framework of covariant canonical transformation theory
in the Hamiltonian picture pioneered by Struckmeier et al.~\citep{struckmeier08}.
That theory is based just on four postulates:
\begin{description}
  \item[Hamilton's Principle] also referred to as the Principle of Least Action, states that the dynamics, i.e.\
  the field equations of motion of a system of classical physical fields must be derived by variation from an action functional.
  \item[Non-Degeneracy] of the Lagrangian is essential for the Legendre transform from the Hamiltonian
  to the Lagrangian picture (and vice versa) to exist, hence to establish the duality transformation of moments and velocities.
  This ensures the applicability of the Hamiltonian canonical transformation theory.
  \item[Diffeomorphism invariance] is required to ensure the invariance of the description of physics---the field equations
  of motion---under chart transitions of the base manifold.
  Hence, the Hamiltonian must be covariant under arbitrary coordinate transformations (diffeomorphisms).
  This is what Einstein had in mind by his Principle of General Relativity \citep{einstein50}.
  \item[Equivalence Principle] means that locally the space-time must be equivalent to an inertial system invariant under Lorentz transformations.
\end{description}
The restriction to four fundamental underlying assumptions is possible as the canonical transformation framework provides
a strong formal guidance to maintain the form of the action principle and hence of the emerging field equations of motion.
Moreover, ambiguities in the form of the dynamics of space-time and its coupling to matter are avoided.
The validity of this approach was proven for ordinary gauge theories and shown to deliver from first principles the correct
Hamiltonian for any SU$(N)$ gauge theory~\citep{StrRei12,struckmeier08,struckmeier17}.
For a dynamical space-time, the approach was extended to a canonical gauge theory of gravity~\citep{struckmeier17a,struckmeier17b}
with scalar (spin-$0$) and vector (spin-$1$) fields as sources for the space-time dynamics.
This paper is an extension of that previous work, which now derives the gravitational coupling of spin-$\nicefrac{1}{2}$ fields.
To this end, an additional structural element is needed for the description of space-time, namely a global
orthonormal (so called tetrad or vierbein) basis attached to every point of the tangent space on the base manifold.
Hence, according to the fourth postulate in the above list, we request the frame of any
observer to be the inertial space, where the metric is globally Minkowskian.
This ``Lorentzian space-time'' is represented by a frame bundle, and the tetrad
field is a global section that pulls back the Minkowski metric to the curved base manifold.
With the inclusion of that frame we have to deal with the additional Lorentz symmetry and combine the
diffeomorphism covariance requirement for the base manifold with the Lorentz covariance of the locally attached inertial frames.
The resulting symmetry group, $\mathrm{Diff}(M)\times\mathrm{SO}(1,3)$, generalizes
the ``affine space'' of the Poincar\'{e} gauge theory~\citep{hehl76,hehl14}.

Throughout the paper we retain the elementary tensor calculus and apply the conventions of~\cite{misner}.
The request for local invariance with respect to both, the Lorentz transformations on the frame bundle
attached to the tangent space, and the diffeomorphism group of chart transitions on the base manifold,
is implemented in Sect.~\ref{sec:complete-cantra} via the choice of a generating function, specifically designed for the underlying symmetry group.
From this point on, the derivation of the diffeomorphism-invariant action is unambiguous and straightforward.
The action, presented in Sect.~\ref{sec:ge-cov-act}, serves as the basis to set up the total closed set of
canonical field equations for the coupled dynamics of fermions an space-time.
In particular, the Dirac equation in curved space-time is worked out.
Regularity of the Dirac Hamiltonian invokes a new length parameter $1/M$ that,
while spurious in the case of non-interacting spinors, becomes a physical parameter
once interaction with space-time or other gauge fields is turned on.
The effective mass of the fermion field acquires an anomalous curvature-dependent mass term,
and novel spin-dependent contributions that couple to the torsion of space-time.
The curvature dependent mass term may have a considerable impact on the physics of dense matter in neutron stars
and around black holes, and also on cosmology~\citep{struckvasak18c,vasak20}.

As all gauge theories merely provide the coupling of the given fields with the gauge fields,
the Hamiltonians describing their free (uncoupled) dynamics must be provided based on physical reasoning.
A particular choice of the Hamiltonian of the free gravitational field---going beyond the version advocated
in Ref.~\cite{struckmeier17a} to accommodate both metric and connection---is finally addressed in Sect.~\ref{sec:free-grav-ham}.
We conclude the paper in Sect.~\ref{sec:conclusions} with a summary and an outlook.
\section{Action principle\label{sec:action_principle}}
\subsection{Hamiltonian action principle in flat space-time}
All Standard Model field theories are based on the Action Principle, which requires that
the information on the dynamical system is encoded in the system's Lagrangian $\LCd$.
The field equations then follow from the extreme of the action $S_0$:
\begin{equation}\label{eq:action-lag}
S_0=\int_V\LCd\,\dd^{4}x,\qquad\delta S_0\stackrel{!}{=}0,
\end{equation}
with $V$ denoting a region of space-time where $\varphi$ and its derivatives are known on the boundary hypersurface $\partial V$.
In the actual context, we assume the field to vanish at infinity.
In a static space-time background, the volume form $\dd^{4}x$ is invariant under the space-time evolution of the fields.
For our purpose of a gauge formalism, we switch to the (covariant) De~Donder-Weyl Hamiltonian $\HCd$~\citep{dedonder30,weyl35}
by means of a \textbf{complete} Legendre transformation.
In analogy to classical mechanics, we have in the simplest case of a scalar field $\varphi$:
\begin{equation*}
H=p\,\dfrac{q}{t}-L\quad\longleftrightarrow\quad\HCd=\pi^{j}\,\pfrac{\varphi}{x^{j}}-\LCd,\quad j=0\ldots 3.
\end{equation*}
The canonical momentum vector $\pi^j(x)$ thus represents the dual of the gradient covector $\partial\varphi/\partial x^j$,
with the Latin indices referring to a Lorentz frame with Minkowski metric $\eta_{ij}$.
The Hamiltonian form of the action of scalar field theories~(\ref{eq:action-lag}) is then the space-time integral:
\begin{equation*}
S_0=\int_V\left[\pi^{j}\,\pfrac{\varphi}{x^{j}}-\HCd\big(\varphi,\pi^i\big)\right]\dd^{4}x.
\end{equation*}
The extreme $\delta S_0\stackrel{!}{=}0$ is encountered exactly if the canonical field equations hold:
\begin{equation}\label{eq:caneq-scalar}
\pfrac{\varphi}{x^{i}}=\pfrac{\HCd}{\pi^{i}},\qquad\pfrac{\pi^{j}}{x^{j}}=-\pfrac{\HCd}{\varphi}.
\end{equation}
We observe that the dependence of $\HCd$ on $\pi^i$ uniquely determines the derivative of the scalar field $\varphi$,
whereas the dependence of $\HCd$ on $\varphi$ merely determines the divergence of the canonical momentum vector $\pi^i$.
This gives rise to a gauge freedom of the canonical momentum vector as any divergence-free vector $p^i$ may be added to $\pi^i$
without violating the canonical field equations~(\ref{eq:caneq-scalar}).
\subsection{Hamiltonian action principle in curvilinear space-time}
For a curvilinear space-time, metric $g_{\mu\nu}(x)$ as well as the volume form $\dd^{4}x$ are no longer invariant.
For the description of spinor fields being the spin-$\nicefrac{1}{2}$ representation of the Lorentz group, it is necessary to introduce tetrads
$e\indices{^i_\mu}(x)$ as new fields representing the geometry of the inertial frames.
The tetrads map the Lorentz frame (Latin indices) with static Minkowski metric $\eta_{ij}$
into the coordinate frame (Greek indices) with a space-time-dependent metric $g_{\mu\nu}(x)$:
\begin{align*}
g_{\mu\nu}&=e\indices{_\mu^i}\,\eta_{ij}\,e\indices{^j_\nu},&e\indices{_\mu^i}\,e\indices{_i^\nu}&=\delta_\mu^\nu,\\
\eta_{ij}&=e\indices{_i^\mu}\,g_{\mu\nu}\,e\indices{^\nu_j},& e\indices{^i_\alpha}\,e\indices{^\alpha_j}&=\delta_j^i,\\
g\equiv\det\left(g_{\mu\nu}\right)&=-{\left(\det e\indices{^i_\mu}\right)}^2\equiv-\varepsilon^2,&\varepsilon&=\sqrt{-g}.
\end{align*}
The invariant volume form is given with $\varepsilon\equiv\det{e\indices{^i_\mu}}$ by $\varepsilon\,\dd^{4}x\equiv\sqrt{-g}\,\dd^{4}x$.
Here, we use the factor $\varepsilon$ to convert the absolute scalar Hamiltonian $\HCd$
into a relative scalar $\tilde{\HCd}=\HCd\varepsilon$ of weight \mbox{$w=1$}.
Correspondingly, the canonical momentum tensors are thus converted into momentum tensor densities---denoted by the tilde---as new dynamical variables:
\begin{equation*}
\tilde{\pi}^{\mu}=\pi^{\mu}\varepsilon,\quad\tilde{k}\indices{_i^{\mu\nu}}=k\indices{_i^{\mu\nu}}\varepsilon,
\end{equation*}
where $\tilde{k}\indices{_i^{\mu\nu}}(x)$ is the tensor density representing the canonical conjugates of the tetrad fields $e\indices{_\mu^i}(x)$.
We then encounter the following form of the action principle which includes the tetrad field to account for the effect of curvilinear geometry:
\begin{equation}\label{eq:scal-action}
S_0=\int_V\left[\tilde{\pi}^{\alpha}\,\pfrac{\varphi}{x^{\alpha}}+\tilde{k}\indices{_j^{\beta\alpha}}\,
\pfrac{e\indices{_\beta^j}}{x^{\alpha}}-\tilde{\HCd}\left(\varphi,\tilde{\pi}^\nu,e\indices{_\mu^i},\tilde{k}\indices{_i^{\mu\nu}}\right)\right]\dd^{4}x.
\end{equation}
Frequently used identities involving the tetrads are:
\begin{equation*}
\pfrac{e\indices{_j^\nu}}{e\indices{_\mu^i}}=-e\indices{_j^\mu}e\indices{_i^\nu},\qquad
\pfrac{\varepsilon}{e\indices{_\mu^i}}=e\indices{_i^\mu}\varepsilon,\qquad
\pfrac{\varepsilon}{x^\nu}=-\varepsilon\,e\indices{^j_\alpha}\pfrac{e\indices{^\alpha_j}}{x^\nu}.
\end{equation*}
\subsection{Klein-Gordon Hamiltonian in curvilinear space-time}
The simplest non-trivial case is given by the Klein-Gordon Hamiltonian for a real scalar field in curvilinear space-time:
\begin{equation*}
\tilde{\HCd}_{\mathrm{KG}}\left(\varphi,\tilde{\pi}^\nu,e\indices{_\mu^i}\right)=
\frac{1}{2\varepsilon}\,\tilde{\pi}^\alpha\,e\indices{_\alpha^i}\,\eta_{ij}\,e\indices{^j_\beta}\,\tilde{\pi}^\beta+\frac{\varepsilon}{2}\,m^2\varphi^2.
\end{equation*}
The field equations follow as
\begin{subequations}
\begin{align}
\pfrac{\varphi}{x^\nu}&=\hphantom{-}\pfrac{\tilde{\HCd}_{\mathrm{KG}}}{\tilde{\pi}^\nu}
=\frac{1}{\dete}\,\tilde{\pi}^\alpha\,e\indices{_\alpha^i}\,\eta_{ij}\,e\indices{^j_\nu}=\pi_\nu\\
\pfrac{\tilde{\pi}^\alpha}{x^\alpha}&=-\pfrac{\tilde{\HCd}_{\mathrm{KG}}}{\varphi}=-\dete\,m^2\,\varphi\\
\pfrac{e\indices{^i_\mu}}{x^\nu}&=\hphantom{-}\pfrac{\tilde{\HCd}_{\mathrm{KG}}}{\tilde{k}\indices{_i^{\mu\nu}}}=0\label{eq:feq-KG-3}\\
\pfrac{\tilde{k}\indices{_i^{\mu\alpha}}}{x^\alpha}&=-\pfrac{\tilde{\HCd}_{\mathrm{KG}}}{e\indices{_\mu^i}}
=-\frac{1}{\dete}\,\tilde{\pi}^\mu\,\eta_{ij}\,e\indices{^j_\beta}\,\tilde{\pi}^\beta\label{eq:feq-KG-4}\\
&\quad+\frac{1}{2}e\indices{_i^\mu}\left(\frac{1}{\dete}\tilde{\pi}^\alpha\,e\indices{_\alpha^n}\,
\eta_{nj}\,e\indices{^j_\beta}\,\tilde{\pi}^\beta-\varepsilon\,m^2\varphi^2\right).\nonumber
\end{align}
\end{subequations}
Solving the first equation for $\tilde{\pi}^\alpha$
\begin{equation*}
\tilde{\pi}^\alpha=\dete\,e\indices{^\alpha_i}\,\eta^{ij}\,e\indices{_j^\beta}\pfrac{\varphi}{x^\beta},
\end{equation*}
the canonical momentum vector can be eliminated from the second equation to yield
\begin{equation*}
\pfrac{}{x^\alpha}\left(\dete\,e\indices{^\alpha_i}\,\eta^{ij}\,e\indices{_j^\beta}\pfrac{\varphi}{x^\beta}\right)+\dete\,m^2\,\varphi=0,
\end{equation*}
which is equivalently expressed in terms of the metric as
\begin{equation*}
g^{\alpha\beta}\ppfrac{\varphi}{x^\beta}{x^\alpha}+\frac{1}{\varepsilon}\pfrac{\varphi}{x^\beta}\pfrac{}{x^\alpha}\left(\dete\,g^{\alpha\beta}\right)+m^2\,\varphi=0.
\end{equation*}
The second term vanishes for a flat metric and thus reproduces the usual Klein-Gordon equation.

In the actual example, the Hamiltonian does not depend on the momentum density $\tilde{k}\indices{_i^{\mu\nu}}$,
which reduces to a Lagrange multiplier in the Lagrangian, i.e.\ in the integrand in the action~(\ref{eq:scal-action}).
Consequently, its conjugate quantity, i.e.\ the metric, is a conserved quantity.
This may change though if the description of the space-time dynamics in the system Hamiltonian is taken into account.

The last canonical equation can be expressed in terms of the first one and the metric energy-momentum tensor density
which in the Hamiltonian representation is the derivative of $\tilde{\HCd}_{\mathrm{KG}}$ with respect to $e\indices{_\mu^j}$:
\begin{align}
\tilde{T}\indices{^\mu_\nu}\equiv\pfrac{\tilde{\HCd}_{\mathrm{KG}}}{e\indices{_\mu^j}}e\indices{_\nu^j}
&=\tilde{\pi}^\mu\,\pfrac{\tilde{\HCd}_{\mathrm{KG}}}{\tilde{\pi}^\nu}-\delta_\nu^\mu\left(
\tilde{\pi}^\alpha\pfrac{\tilde{\HCd}_{\mathrm{KG}}}{\tilde{\pi}^\alpha}-\tilde{\HCd}_{\mathrm{KG}}\right)\nonumber\\
&=\tilde{\theta}\indices{^\mu_\nu}.\label{def:TKG}
\end{align}
The right-hand side is exactly the Hamiltonian form of the canonical energy-momentum tensor density,
which happens to agree with the metric one for the Klein-Gordon system.
Regrouping the terms yields the Hamiltonian representation of the identity~\citep{struckmeier18a} for the scalar density function
$\tilde{\HCd}_{\mathrm{KG}}\left(\varphi,\tilde{\pi}^\nu,e\indices{_\mu^i}\right)$:
\begin{equation*}
\pfrac{\tilde{\HCd}_{\mathrm{KG}}}{e\indices{_\mu^j}}e\indices{_\nu^j}-
\pfrac{\tilde{\HCd}_{\mathrm{KG}}}{\tilde{\pi}^\nu}\,\tilde{\pi}^\mu+
\delta_\nu^\mu\left(
\pfrac{\tilde{\HCd}_{\mathrm{KG}}}{\tilde{\pi}^\alpha}\tilde{\pi}^\alpha
-\tilde{\HCd}_{\mathrm{KG}}\right)\equiv0.
\end{equation*}
\subsection{Dirac Hamiltonian in curvilinear space-time}
The regularized Dirac Lagrangian density for spinors in curvilinear space-time is given by~\citep{gasiorowicz66,struckmeier08,struckvasak18c}:
\begin{align}
\tilde{\LCd}_{\mathrm{D}}&=\frac{\rmi\dete}{3M}\left(\pfrac{\psibar}{x^\alpha}\,e\indices{^\alpha_k}-
\frac{\rmi M}{2}\psibar\,\gamma_k\right)\sigma^{kj}
\left(e\indices{_j^\beta}\,\pfrac{\psi}{x^\beta}+\frac{\rmi M}{2}\gamma_j\,\psi\right)\nonumber\\
&\quad-\left(m-M\right)\psibar\psi\,\dete,\label{eq:ld-dirac}
\end{align}
with $m$ the usual mass of the Dirac particle and $M$ a free parameter of mass dimension.
Due to the quadratic ``velocity'' dependence of~(\ref{eq:ld-dirac}), the corresponding
covariant Hamiltonian as obtained via the Legendre transformation is
\begin{align*}
\tilde{\HCd}_{\mathrm{D}}\left(\psi,\tilde{\kappabar}^\nu,\psibar,\tilde{\kappa}^\nu,e\indices{^\nu_k}\right)
&=\tilde{\kappabar}^\alpha\pfrac{\psi}{x^\alpha}+\pfrac{\psibar}{x^\alpha}\tilde{\kappa}^\alpha\\
&\quad-\tilde{\LCd}_{\mathrm{D}}\left(\psi,\partial_\nu\psi,\psibar,\partial_\nu\psibar,e\indices{^\nu_k}\right)
\end{align*}
with the canonical momenta $\tilde{\kappabar}^\nu$ and $\tilde{\kappa}^\nu$ defined by:
\begin{equation*}
\tilde{\kappabar}^\nu=\pfrac{\tilde{\LCd}_{\mathrm{D}}}{\left(\pfrac{\psi}{x^\nu}\right)},\qquad
\tilde{\kappa}^\nu=\pfrac{\tilde{\LCd}_{\mathrm{D}}}{\left(\pfrac{\psibar}{x^\nu}\right)}.
\end{equation*}
With Eq.~(\ref{eq:ld-dirac}) the Dirac Hamiltonian density $\tilde{\HCd}_{\mathrm{D}}$ then follows as
\begin{equation}\label{hd-dirac}
\tilde{\HCd}_{\mathrm{D}}=\frac{3M}{\rmi\dete}
\left(\tilde{\kappabar}^{\alpha}e\indices{_\alpha^k}-\frac{\rmi\dete}{2}\bar{\psi}\gamma^{k}\right)
\tau_{kj}\left(e\indices{^j_\beta}\tilde{\kappa}^{\beta}+\frac{\rmi\dete}{2}\gamma^{j}\psi\right)+m\,\bar{\psi}\psi\dete,
\end{equation}
or, equivalently in expanded form,
\begin{align}
\tilde{\HCd}_{\mathrm{D}}&=\frac{\rmi M}{2}\left(\psibar\,\gamma_j\,e\indices{^j_\beta}\,\tilde{\kappa}^\beta
-\frac{6}{\dete}\,\tilde{\kappabar}^\alpha\,e\indices{_\alpha^k}\,\tau_{kj}\,e\indices{^j_\beta}\,\tilde{\kappa}^\beta
-\tilde{\kappabar}^\alpha\,e\indices{_\alpha^k}\,\gamma_k\,\psi\right)\nonumber\\
&\quad+\left(m-M\right)\psibar\psi\,\dete.
\label{eq:hd-dirac}
\end{align}
$\tau_{kj}$ is the inverse of the commutator $\sigma^{jk}$ of the Dirac matrices:
\begin{align*}
\sigma^{jk}&\equiv\frac{\rmi}{2}\left(\gamma^j\gamma^k-\gamma^k\gamma^j\right),&
\eta^{jk}\Eins&=\frac{1}{2}\left(\gamma^j\gamma^k+\gamma^k\gamma^j\right)\\
\tau_{kj}&\equiv\frac{\rmi}{6}\left(\gamma_{k}\gamma_{j}+3\gamma_{j}\gamma_{k}\right),&
\tau_{ik}\sigma^{kj}&=\delta_{i}^{j}\,\Eins.
\end{align*}
Here $\eta_{ik}$ is the Minkowski metric, and $\Eins$ the unit matrix in spinor space.
These definitions imply the identities:
\begin{equation}\label{eq:tau-identity}
\gamma_k\,\sigma^{kj}\equiv\sigma^{jk}\,\gamma_k\equiv 3\rmi\,\gamma^j,\qquad
\gamma^{k}\tau_{kj}\equiv\tau_{jk}\gamma^{k}\equiv\frac{1}{3\rmi}\,\gamma_{j}.
\end{equation}
Setting up the covariant canonical equations for the Hamiltonian~(\ref{eq:hd-dirac}), gives:
\begin{subequations}\label{eq:can-dirac-1-4}
\begin{align}
\pfrac{\psi}{x^\nu}&=\hphantom{-}\pfrac{\tilde{\HCd}_{\mathrm{D}}}{\tilde{\kappabar}^\nu}
=-\frac{\rmi M}{2}e\indices{_\nu^k}\left(\gamma_k\,\psi+\frac{6}{\dete}\,
\tau_{kj}\,e\indices{^j_\beta}\,\tilde{\kappa}^\beta\right)\label{eq:can-dirac1}\\
\pfrac{\tilde{\kappa}^\alpha}{x^\alpha}&=-\pfrac{\tilde{\HCd}_{\mathrm{D}}}{\psibar}=
-\frac{\rmi M}{2}\gamma_j\,e\indices{^j_\beta}\,\tilde{\kappa}^\beta-\left(m-M\right)\psi\,\dete\label{eq:can-dirac2}\\
\pfrac{\psibar}{x^{\nu}}&=\hphantom{-}\pfrac{\tilde{\HCd}_{\mathrm{D}}}{\tilde{\kappa}^{\nu}}
=\hphantom{-}\frac{\rmi M}{2}\left(\psibar\,\gamma_j-\frac{6}{\dete}\,\tilde{\kappabar}^\alpha\,
e\indices{_\alpha^k}\,\tau_{kj}\right)e\indices{^j_\nu}\label{eq:can-dirac3}\\
\pfrac{\tilde{\kappabar}^{\alpha}}{x^{\alpha}}&=-\pfrac{\tilde{\HCd}_{\mathrm{D}}}{\psi}
=\hphantom{-}\frac{\rmi M}{2}\tilde{\kappabar}^\alpha\,e\indices{_\alpha^k}\,\gamma_k-\left(m-M\right)\psibar\,\dete.\label{eq:can-dirac4}
\end{align}
\end{subequations}
Equation~(\ref{eq:can-dirac1}) can be solved for $\tilde{\kappa}^\mu$ and Eq.~(\ref{eq:can-dirac3}) for $\tilde{\kappabar}^\mu$:
\begin{subequations}\label{eq:dirac-momenta}
\begin{align}
\tilde{\kappa}^\mu&=e\indices{^\mu_n}\left(\frac{\rmi}{3M}\sigma^{nm}\,e\indices{_m^\beta}\,
\pfrac{\psi}{x^\beta}-\frac{\rmi}{2}\gamma^n\,\psi\right)\dete\label{eq:can-dirac1a}\\
\tilde{\kappabar}^\mu&=\left(\frac{\rmi}{3M}\pfrac{\psibar}{x^\beta}\,e\indices{^\beta_m}\,\sigma^{mn}
+\frac{\rmi}{2}\psibar\,\gamma^n\right)e\indices{_n^\mu}\,\dete.\label{eq:can-dirac3a}
\end{align}
\end{subequations}
Inserting Eq.~(\ref{eq:can-dirac1a}) into Eq.~(\ref{eq:can-dirac2})
yields the generalized Dirac equation in curvilinear space-time:
\begin{align}
0&=\rmi\gamma^{k}\,e\indices{_k^\alpha}\pfrac{\psi}{x^\alpha}-m\,\psi+\frac{\rmi}{2}\gamma^k\left(\pfrac{e\indices{_k^\alpha}}{x^\alpha}
-e\indices{_k^\alpha}\,\pfrac{e\indices{^\xi_i}}{x^\alpha}e\indices{^i_\xi}\right)\psi\nonumber\\
&\quad-\frac{\rmi\sigma^{kj}}{3M}\left(
\pfrac{e\indices{^\alpha_k}}{x^\alpha}e\indices{_j^\beta}+e\indices{^\alpha_k}\pfrac{e\indices{_j^\beta}}{x^\alpha}-
e\indices{^\alpha_k}e\indices{_j^\beta}\pfrac{e\indices{^\xi_i}}{x^\alpha}e\indices{^i_\xi}\right)\pfrac{\psi}{x^\beta}.
\label{eq:dirac-equation}
\end{align}
It obviously reduces to the usual Dirac equation in a flat space-time geometry where all
derivatives of the tetrads vanish.
Inserting Eq.~(\ref{eq:can-dirac3a}) into Eq.~(\ref{eq:can-dirac4})
yields the generalized Dirac equation for the adjoint spinor $\psibar$.
We remark that in the case of the Hamiltonian description the term quadratic in the canonical momenta
$\tilde{\kappabar}^\alpha$ and $\tilde{\kappa}^\beta$ in~(\ref{eq:hd-dirac}) is mandatory in this formulation as
otherwise---according to Eqs.~(\ref{eq:can-dirac-1-4})---no correlation
would exist between canonical momenta and ``velocities'', i.e.\ the space-time derivatives of the spinors.

The metric energy-momentum tensor density $\tilde{T}\indices{^\mu_\nu}$ of the Dirac system is now, in analogy to Eq.~(\ref{def:TKG}),
\begin{align}
\tilde{T}\indices{^\mu_\nu}=\pfrac{\tilde{\HCd}_{\mathrm{D}}}{e\indices{_\mu^j}}e\indices{_\nu^j}
&=\frac{\rmi M}{2}\left[\left(\psibar\,\gamma_j-\frac{6}{\dete}\tilde{\kappabar}^\alpha\,e\indices{_\alpha^k}
\tau_{kj}\right)e\indices{^j_\nu}\,\tilde{\kappa}^\mu\right.\nonumber\\
&\left.\qquad\quad\mbox{}-\tilde{\kappabar}^\mu\,e\indices{_\nu^j}\left(\gamma_j\,\psi
+\frac{6}{\dete}\tau_{jk}\,e\indices{^k_\beta}\,\tilde{\kappa}^\beta\right)\right]\nonumber\\
&+\delta_\nu^\mu\!\left[\frac{3\rmi M}{\dete}\tilde{\kappabar}^\alpha e\indices{_\alpha^k}
\tau_{kj}\,e\indices{^j_\beta}\tilde{\kappa}^\beta+\left(m-M\right)\psibar\psi\,\dete\right]\!.
\label{eq:emt-dirac}
\end{align}
The canonical energy-momentum tensor density $\tilde{\theta}\indices{^\mu_\nu}$ of the Dirac system, defined in the Hamiltonian representation by
\begin{equation*}
\tilde{\theta}\indices{^\mu_\nu}=\pfrac{\tilde{\HCd}_{\mathrm{D}}}{\tilde{\kappa}^\nu}\tilde{\kappa}^\mu
+\tilde{\kappabar}^\mu\pfrac{\tilde{\HCd}_{\mathrm{D}}}{\tilde{\kappabar}^\nu}
-\delta_\nu^\mu\left(\tilde{\kappabar}^\beta\,\pfrac{\tilde{\HCd}_{\mathrm{D}}}{\tilde{\kappabar}^\beta}
+\pfrac{\tilde{\HCd}_{\mathrm{D}}}{\tilde{\kappa}^\beta}\,\tilde{\kappa}^\beta-\tilde{\HCd}_{\mathrm{D}}\right)
\end{equation*}
can be shown by means of the canonical equations~(\ref{eq:can-dirac-1-4}) to
coincide with the metric energy-momentum tensor: $\tilde{\theta}\indices{^\mu_\nu}\equiv\tilde{T}\indices{^\mu_\nu}$.
\subsection{Requirement of form-invariance of the action under diffeomorphisms}
Implementing the postulate that the equations of physics should be independent of the
reference frame means to request form-invariance of the action under diffeomorphisms.
An inspection of the action integral~(\ref{eq:scal-action}) already shows that this demand is in general not met in a curvilinear space-time
as the derivatives of non-scalar quantities do not transform covariantly under chart transitions of the space-time manifold.
This applies, for instance, to the tetrads $e\indices{^j_\mu}$, which represent generalized tensors that
reside in both the general coordinate space (Greek index) and in the local inertial frame (Latin index).
Its tensor transformation rule $e\indices{^j_\mu}(x)\mapsto E\indices{^I_\nu}(X)$ under a chart transition $x\mapsto X$,
and an arbitrary Lorentz transformation in the local inertial space, is given by:
\begin{equation}\label{def:tetradtransform}
E\indices{^I_\nu}(X)=\Lambda\indices{^I_j}(x)\,e\indices{^j_\beta}(x)\pfrac{x^\beta}{X^\nu}.
\end{equation}
$\Lambda\indices{^I_j}(x)$ denotes the skew-symmetric matrix of local (orthochronous) Lorentz transformations
in the inertial frame, $\Lambda\indices{_I_j}=-\Lambda\indices{_j_I}$.
Here and in the following capital letters denote transformed fields or indices in a transformed inertial frame.
Thus
\begin{equation*}
\Lambda\indices{^I_k}\,\Lambda\indices{^k_J}=\Lambda\indices{_J^k}\,\Lambda\indices{_k^I}=\delta_J^I
\quad\Leftrightarrow\quad
\Lambda\indices{^i_K}\,\Lambda\indices{^K_j}=\Lambda\indices{_j^K}\,\Lambda\indices{_K^i}=\delta_j^i.
\end{equation*}
The derivative of the tetrad $e\indices{^j_\mu}$ in the action integral~(\ref{eq:scal-action}) does not transform as a tensor in a general space-time geometry:
\begin{equation}\label{def:tetraddertransform}
\pfrac{E\indices{^J_\nu}}{X^\xi}=\Bigg(\pfrac{\Lambda\indices{^J_j}}{x^\alpha}e\indices{^j_\beta}+\Lambda\indices{^J_j}\pfrac{e\indices{^j_\beta}}{x^\alpha}
\Bigg)\pfrac{x^\alpha}{X^\xi}\pfrac{x^\beta}{X^\nu}+\Lambda\indices{^J_j}e\indices{^j_\beta}\ppfrac{x^\beta}{X^\nu}{X^\xi}.
\end{equation}
The last term spoils the tensor transformation property in a curved space-time, as the second derivatives of $x^\beta(X)$ do not identically vanish.
As a consequence, the action~(\ref{eq:scal-action}) is not diffeomorphism-invariant.
In order to render actions invariant, one must proceed as follows:
\begin{itemize}
\item The second partial derivatives must be compensated away by means of formally introducing an appropriate gauge field.
This provides the coupling of the given fields to the gauge field, and converts partial into covariant derivatives.
\item The description of the gauge field dynamics must be part of the final action integral in order to end up with a closed dynamical system,
hence a system which does not contain external fields.
This is achieved by postulating the corresponding Hamiltonian of the free (uncoupled) gauge field dynamics.
\end{itemize}
For all systems whose dynamics are derived from an action principle, any transformation
must be canonical in order to maintain the general form of the canonical field equations.
Thus, in particular gauge theories are in the end most easily formulated within the canonical transformation
framework as non-canonical and hence unphysical transformations are excluded at the outset.
\section{Canonical transformation framework}
\subsection{Canonical transformation formalism for a scalar field in a curvilinear space-time}
A scalar Hamiltonian that depends on a set of fields and dynamical tetrads and is invariant under a global Lorentz transformation
will not in general be invariant under a combined arbitrary diffeomorphisms $x\mapsto X(x)$ at a given point of the base manifold,
and local Lorentz transformations of the frames attached to that point.
Considering for illustration the transformations of a scalar field, $\varphi(x)\mapsto\Phi(X)$, and of the tetrads,
$e\indices{^i_\alpha}(x)\mapsto E\indices{^I_\beta}(X)$, the requested invariance of the equations of motion means explicitly:
\begin{align}
&\delta S_0\!=\!\delta\!\!\int_V\!\!\Bigg[\tilde{\bar{\pi}}^\alpha\pfrac{\varphi}{x^{\alpha}}+
\tilde{k}\indices{_i^{\beta\alpha}}\pfrac{e\indices{_\beta^i}}{x^{\alpha}}-
\tilde{\HCd}\left(\varphi,\tilde{\pi}^\nu,e\indices{^i_\mu},\tilde{k}\indices{_i^{\mu\nu}},x\right)\!\Bigg]\!\dd^4x\nonumber\\
&\must\!\delta\!\!\!\int_{V^\prime}\!\!\Bigg[\tilde{\bar{\Pi}}^\alpha\!\pfrac{\Phi}{X^{\alpha}}+
\tilde{K}\indices{_I^{\beta\alpha}}\pfrac{E\indices{^I_\beta}}{X^{\alpha}}-
\tilde{\HCd}^{\prime}\!\left(\Phi,\tilde{\Pi}^\nu\!,\!E\indices{^I_\mu},\!\tilde{K}\indices{_I^{\mu\nu}}\!,\!X\right)\!\!\Bigg]\!\dd^4X\!.
\label{eq:action-condistion}
\end{align}
As the actions are to be varied in order to derive the canonical field equations, the integrands
of Eq.~(\ref{eq:action-condistion}) may differ by the divergence of an arbitrary vector function $\tilde{\FCd}_{1}^{\mu}$.
Such a term does not contribute to the variation of $S_0$ by virtue of Gauss' law for $\tilde{\FCd}_1^\alpha=\FCd_1^\alpha\sqrt{-g}$,
i.e.\ the product of the absolute vector field $\FCd_1^\alpha$ with the scalar field $\sqrt{-g}$,
\begin{equation*}
\delta\int_{V}\pfrac{\tilde{\FCd}_1^\alpha}{x^\alpha}\,\dd^4x=\delta\oint_{\partial V}\tilde{\FCd}_1^\alpha\,\dd S_\alpha\,\must\,0,
\end{equation*}
as the variation is supposed to vanish on the boundary $\partial V$.
With the volume form $\dd^{4}x$ transforming as a relative scalar of weight $w=-1$,
\begin{equation}\label{eq:trans-volumeform}
\dd^{4}X=\pfrac{\left(X^{0},\ldots,X^{3}\right)}{\left(x^{0},\ldots,x^{3}\right)}\dd^{4}x=
\detpartial{X}{x}\dd^{4}x=\frac{1}{\detpartial{x}{X}}\dd^{4}x,
\end{equation}
the integrands in Eq.~(\ref{eq:action-condistion}) must satisfy the equation:
\begin{align}
&\tilde{\pi}^\alpha\!\pfrac{\varphi}{x^\alpha}\!+\!\tilde{k}\indices{_i^{\mu\nu}}\pfrac{e\indices{^i_\mu}}{x^\nu}-
\tilde{\HCd}-\Bigg(\tilde{\Pi}^\nu\pfrac{\Phi}{X^\nu}\!+\!\tilde{K}\indices{_I^{\mu\nu}}\pfrac{E\indices{^I_\mu}}{X^\nu}-
\tilde{\HCd}^\prime\Bigg)\!\detpartial{X}{x}\nonumber\\
&=\!\pfrac{\tilde{\FCd}_1^\nu}{\varphi}\pfrac{\varphi}{x^\nu}+\pfrac{\tilde{\FCd}_1^\beta}{\Phi}\pfrac{X^\nu}{x^\beta}\pfrac{\Phi}{X^\nu}+
\pfrac{\tilde{\FCd}_1^\nu}{e\indices{^i_\mu}}\pfrac{e\indices{^i_\mu}}{x^\nu}+
\pfrac{\tilde{\FCd}_1^\beta}{E\indices{^I_\mu}}\pfrac{X^\nu}{x^\beta}\pfrac{E\indices{^I_\mu}}{X^\nu}\nonumber\\
&\quad+\left.\pfrac{\tilde{\FCd}_1^\alpha}{x^\alpha}\right|_{\text{expl}}.
\label{actionintegrand-ham}
\end{align}
For the particular choice $\tilde{\FCd}_1^\nu=\tilde{\FCd}_1^\nu\big(\varphi,\Phi,e\indices{^i_\mu},E\indices{^I_\mu},x\big)$, we can compare the
coefficients of the partial derivatives of the fields and thereby identify the following transformation rules for the fields:
\begin{subequations}\label{eq:f1-rules}
\begin{align}
\tilde{\pi}^\nu &= \pfrac{\tilde{\FCd}_1^\nu}{\varphi}, &
\tilde{\Pi}^\nu &= -\pfrac{\tilde{\FCd}_1^\beta}{\Phi}\pfrac{X^\nu}{x^\beta}\detpartial{x}{X}\\
\tilde{k}\indices{_i^{\mu\nu}} &= \pfrac{\tilde{\FCd}_1^\nu}{e\indices{^i_\mu}}, &
\tilde{K}\indices{_I^{\mu\nu}} &=-\pfrac{\tilde{\FCd}_1^\beta}{E\indices{^I_\mu}}\pfrac{X^\nu}{x^\beta}\detpartial{x}{X}.
\end{align}
\end{subequations}
The transformation rule for the Hamiltonians involves the possible explicit dependence of $\tilde{\FCd}_1^\nu(x)$ on $x$:
\begin{equation}
\tilde{\HCd}^\prime\detpartial{X}{x}=\tilde{\HCd}+\left.\pfrac{\tilde{\FCd}_1^\alpha}{x^\alpha}\right|_{\text{expl}}.
\end{equation}
A canonical transformation is also generated by a vector density $\tilde{\FCd}_3^\nu$~\citep{struckmeier08}, defined as a function of the momenta
$\tilde{\pi}^\nu$ and $\tilde{k}\indices{_i^{\mu\nu}}$ in place of the fields $\varphi$ and $e\indices{^{i}_{\mu}}$ in $\tilde{\FCd}_1^\nu$.
It is defined as the Legendre transformation of $\tilde{\FCd}_1^\nu$:
\begin{equation*}
\tilde{\FCd}_3^\nu\!\left(\tilde{\pi}^\nu,\Phi,\tilde{k}\indices{_i^{\mu\nu}},E\indices{^{I}_{\mu}},\!x\right)
\!=\!\tilde{\FCd}_1^\nu\!\left(\varphi,\Phi,e\indices{^{i}_{\mu}},E\indices{^{I}_{\mu}},x\right)
-\tilde{\pi}^{\nu}\,\varphi-\tilde{k}\indices{_i^{\mu\nu}}\,e\indices{^{i}_{\mu}}.
\end{equation*}
One then obtains the field transformation rules:
\begin{subequations}\label{eq:f3-rules}
\begin{align}
\delta^\nu_\alpha\,\varphi&=-\pfrac{\tilde{\FCd}_3^\nu}{\tilde{\pi}^\alpha}&
\tilde{\Pi}^\nu&=-\pfrac{\tilde{\FCd}_3^\alpha}{\Phi}\,\pfrac{X^\nu}{x^\alpha}\,\detpartial{x}{X}\\
\delta^\nu_\alpha\,e\indices{^{i}_{\mu}}&=-\pfrac{\tilde{\FCd}_3^\nu}{\tilde{k}\indices{_i^{\mu\alpha}}}&
\tilde{K}\indices{_I^{\mu\nu}}&=-\pfrac{\tilde{\FCd}_3^\alpha}{E\indices{^{I}_{\mu}}}\,\pfrac{X^\nu}{x^\alpha}\,\detpartial{x}{X}
\end{align}
and the similar rule for the Hamiltonians:
\begin{equation}\label{eq:f3derivative}
\tilde{\HCd}^\prime\detpartial{X}{x}=\tilde{\HCd}+\left.\pfrac{\tilde{\FCd}_3^\nu}{x^\nu}\right|_{\text{expl}}.
\end{equation}
\end{subequations}
The untransformed fields are thus correlated to the negative derivatives of the
generating function $\tilde{\FCd}_3^\nu$ with respect to the untransformed conjugate momentum fields.
Furthermore, the transformed conjugate momentum fields are given by the negative derivatives of the
generating function $\tilde{\FCd}_3^\nu$ with respect to the transformed fields times the transition factors for tensor densities
\begin{equation*}
\tilde{\Pi}^\nu(X)=\tilde{\pi}^\alpha(x)\,\pfrac{X^\nu}{x^\alpha}\,\detpartial{x}{X}.
\end{equation*}
This scheme applies as well for all other types of tensor fields with their respective conjugate momentum fields which constitute tensor densities.
\subsection{Diffeomorphism-invariance of a scalar field action integral induced by a gauge field}
In the next step the particular generating function is defined for a canonical transformation
that provides combined Lorentz and chart transformations, while leaving the scalar field $\varphi$ unchanged:
\begin{equation} \label{def:F3_Lorentzandchart}
\tilde{\FCd}_3^\nu (\Phi,\tilde{\pi}^\nu, E\indices{^{I}_{\mu}}, \tilde{k}\indices{_i^{\mu\nu}})
=-\tilde{\pi}^\nu\,\Phi-\tilde{k}\indices{_i^{\beta\nu}}\,\Lambda\indices{^i_I}\,E\indices{^{I}_{\alpha}} \, \pfrac{X^\alpha}{x^\beta}.
\end{equation}
The particular transformation rules~(\ref{eq:f3-rules}) follow as
\begin{subequations}
\begin{align}
\varphi&=\Phi,&\tilde{\Pi}^\nu&=\tilde{\pi}^\alpha \pfrac{X^\nu}{x^\alpha}\detpartial{x}{X}
\label{F3derivative211}\\
e\indices{^{i}_{\mu}}&=\Lambda\indices{^i_I}\,E\indices{^{I}_{\alpha}}\,\pfrac{X^\alpha}{x^\mu},&
\tilde{K}\indices{_I^{\mu\nu}}\,&=\tilde{k}\indices{_i^{\beta\alpha}}\Lambda\indices{^i_I}\,
\pfrac{X^\mu}{x^\beta}\,\pfrac{X^\nu}{x^\alpha}\,\detpartial{x}{X},
\label{F3derivative213}
\end{align}
\end{subequations}
which recover the proper transformation rules for the fields and their conjugates,
the latter transforming as relative tensors of weight $w=1$, i.e., as tensor densities.

The set of transformation rules is completed by the rule for the Hamiltonian density from Eq.~(\ref{eq:f3derivative}),
which follows from the explicitly space-time-dependent coefficients of the generating function~(\ref{def:F3_Lorentzandchart})
\begin{align}
\left.\pfrac{\tilde{\FCd}_3^\nu}{x^\nu} \right|_{\text{expl}}&=
-\tilde{k}\indices{_i^{\beta\nu}}\pfrac{}{x^\nu}\left(\Lambda\indices{^i_I}\pfrac{X^\alpha}{x^\beta}\right)E\indices{^{I}_{\alpha}}\label{HLorentzandchart}\\
&=-\tilde{k}\indices{_i^{[\beta\nu]}}\pfrac{\Lambda\indices{^i_I}}{x^\nu}\pfrac{X^\alpha}{x^\beta}E\indices{^{I}_{\alpha}}-
\tilde{k}\indices{_i^{(\beta\nu)}}\pfrac{}{x^\nu}\!\left(\Lambda\indices{^i_I}\pfrac{X^\alpha}{x^\beta}\right)\!E\indices{^{I}_{\alpha}}.\nonumber
\end{align}
In the last line, the right-hand side of~(\ref{HLorentzandchart}) is split into the skew-symmetric and the
symmetric contributions of $\tilde{k}\indices{_i^{\beta\nu}}$ in $\beta$ and $\nu$, considering that the second
derivative term of $X^\alpha$ does not contribute to the skew-symmetric part of $\tilde{k}\indices{_i^{\nu\beta}}$.

The $x^\nu$-derivative term in~(\ref{HLorentzandchart}) is equivalently expressed
in terms of the derivative of the transformation rule~(\ref{F3derivative213}) for the tetrad $e\indices{^{i}_{\beta}}$:
\begin{equation}\label{deri-F3derivative214}
\pfrac{}{x^\nu}\left(\Lambda\indices{^i_I}\pfrac{X^\alpha}{x^\beta}\right)E\indices{^{I}_{\alpha}}
=\pfrac{e\indices{^{i}_{\beta}}}{x^\nu}-\Lambda\indices{^i_I}\pfrac{E\indices{^{I}_{\alpha}}}{X^\xi}\pfrac{X^\xi}{x^\nu}\pfrac{X^\alpha}{x^\beta}.
\end{equation}
Inserting Eq.~(\ref{deri-F3derivative214}) into the transformation rule~(\ref{HLorentzandchart}) yields:
\begin{align*}
\left.\pfrac{\tilde{\FCd}_3^\nu}{x^\nu} \right|_{\text{expl}}\!\!\!\!
&=-\tilde{k}\indices{_i^{[\beta\nu]}}\pfrac{\Lambda\indices{^i_I}}{x^\nu}\pfrac{X^\alpha}{x^\beta}E\indices{^{I}_{\alpha}}\\
&\quad-\tilde{k}\indices{_i^{(\beta\nu)}}\Bigg(\pfrac{e\indices{^{i}_{\beta}}}{x^\nu}-
\Lambda\indices{^i_I}\pfrac{E\indices{^{I}_{\alpha}}}{X^\xi}\pfrac{X^\xi}{x^\nu}\pfrac{X^\alpha}{x^\beta}\Bigg)\\
&=\tilde{k}\indices{_i^{[\mu\nu]}}\Lambda\indices{^i_I}\pfrac{\Lambda\indices{^I_j}}{x^\nu} e\indices{^{j}_{\mu}}
-\tilde{k}\indices{_i^{(\mu\nu)}}\pfrac{e\indices{^{i}_{\mu}}}{x^\nu}
+\tilde{K}\indices{_I^{(\mu\nu)}} \pfrac{E\indices{^{I}_{\mu}}}{X^\nu}\detpartial{X}{x}\!,
\end{align*}
where in the last equation the transformation rules~(\ref{F3derivative213}) were inserted in the first and the third term.
Plugging this into the condition for the action functionals, the derivative terms of the tetrads can be
merged with corresponding derivatives originating from the Legendre transformation in Eq.~(\ref{eq:action-condistion}) to give
the modified action functionals
\begin{align}
&\delta\int_{V^\prime}\left[\tilde{\Pi}^\nu\pfrac{\Phi}{X^{\nu}}+
\frac{1}{2}\tilde{K}\indices{_I^{\mu\nu}}\left(\pfrac{E\indices{^{I}_{\mu}}}{X^{\nu}}-
\pfrac{E\indices{^{I}_{\nu}}}{X^\mu}\right)-\tilde{\HCd}^\prime\right]\detpartial{X}{x}\dd^{4}x\nonumber\\
&=\delta\int_{V}\Bigg[\tilde{\pi}^\nu\pfrac{\varphi}{x^{\nu}}+
\frac{1}{2}\tilde{k}\indices{_i^{\mu\nu}}\left(\pfrac{e\indices{^{i}_{\mu}}}{x^{\nu}}-\pfrac{e\indices{^{i}_{\nu}}}{x^\mu}\right)-\tilde{\HCd}
\nonumber\\&\qquad\qquad
+\tilde{k}\indices{_i^{[\mu\nu]}}\Lambda\indices{^i_I}\pfrac{\Lambda\indices{^I_j}}{x^\nu}e\indices{^{j}_{\mu}}\Bigg]\,\dd^4x.
\label{F3derivative12a}
\end{align}
Owing to the last term on the right-hand side of Eq.~(\ref{F3derivative12a}), the actions are no longer
form-invariant for space-time-dependent Lorentz transformation coefficients $\Lambda\indices{^I_j}(x)$.
The only way to re-establish the form invariance of the actions is to amend the integrands
by gauge Hamiltonians whose transformation rule absorbs the symmetry-breaking term:
\begin{align*}
&\quad\delta\int_{V\prime}\left(\tilde{\Pi}^\nu\pfrac{\Phi}{X^{\nu}}
+\tilde{K}\indices{_I^{[\mu\nu]}}\pfrac{E\indices{^{I}_{\mu}}}{X^{\nu}}
-\tilde{\HCd}^\prime-\tilde{\HCd}_{\mathrm{Gau}_1}^{\prime}\right)\detpartial{X}{x}\dd^4x\nonumber\\
&=\delta\int_V\Bigg(\tilde{\pi}^\nu\pfrac{\varphi}{x^{\nu}}+\tilde{k}\indices{_i^{[\mu\nu]}}\pfrac{e\indices{^{i}_{\mu}}}{x^{\nu}}
-\tilde{\HCd}-\tilde{\HCd}_{\mathrm{Gau}_1}\Bigg)\,\dd^4x.
\end{align*}
This entails the following transformation requirement for the gauge Hamiltonians:
\begin{equation}\label{eq:inv-cond}
\tilde{\HCd}_{\mathrm{Gau}_1}^{\prime}\,\detpartial{X}{x}-\tilde{\HCd}_{\mathrm{Gau}_1}=
\tilde{k}\indices{_i^{[\mu\nu]}}\Lambda\indices{^i_I}\pfrac{\Lambda\indices{^I_j}}{x^\nu}\,e\indices{^j_\mu}.
\end{equation}
The  gauge Hamiltonian $\tilde{\HCd}_{\mathrm{Gau}_1}$ must be devised in the way that the external index structure of
the coefficient expression $\Lambda\indices{^i_I}\partial\Lambda\indices{^I_j}/\partial x^\nu$
in~(\ref{F3derivative12a}) is matched by a gauge field $\ho\indices{^i_{j\nu}}$.
Its obvious form is (with the negative sign chosen for later convenience)
\begin{equation}\label{g-ham1}
\tilde{\HCd}_{\mathrm{Gau}_1}=-\tilde{k}\indices{_i^{[\mu\nu]}}\,\ho\indices{^i_{j\nu}}\,e\indices{^j_\mu},
\end{equation}
which is required to satisfy both, form-invariance in terms of the transformed gauge field $\HO\indices{^I_{J\nu}}$,
and Eq.~(\ref{eq:inv-cond}) under the transformation in question.
Only then can the above mentioned form invariance of the action integrals be re-established.
Hence:
\begin{equation}\label{g-ham2}
\tilde{\HCd}_{\mathrm{Gau}_1}^{\prime}=-\tilde{K}\indices{_I^{[\mu\nu]}}\,\HO\indices{^I_{J\nu}}\,E\indices{^J_\mu}.
\end{equation}
As the gauge field $\ho\indices{^i_{j\nu}}$ now replaces the coefficient expression $\Lambda\indices{^i_I}\partial\Lambda\indices{^I_j}/\partial x^\nu$,
the skew-symmetry of the coefficient matrix $\Lambda\indices{_j_I}=-\Lambda\indices{_I_j}$
of the local Lorentz transformation induces the gauge field $\ho\indices{_i_{j\nu}}$ to be skew-symmetric in $i,j$:
\begin{align*}
\ho\indices{_i_{j\nu}}\leftrightarrow\Lambda\indices{_i_I}\pfrac{\Lambda\indices{^I_j}}{x^\nu}&=-\Lambda\indices{_I_i}\pfrac{\Lambda\indices{^I_j}}{x^\nu}
=\pfrac{\Lambda\indices{_I_i}}{x^\nu}\Lambda\indices{^I_j}=\pfrac{\Lambda\indices{^I_i}}{x^\nu}\Lambda\indices{_I_j}\\
&=-\Lambda\indices{_j_I}\pfrac{\Lambda\indices{^I_i}}{x^\nu}\leftrightarrow-\ho\indices{_j_{i\nu}}.
\end{align*}
Here the fact has been used that the metric of the inertial frames, $\eta_{IJ}$, is by definition  globally constant, hence $\partial\eta_{IJ}/\partial x^\nu\equiv0$.

Now the ensuing transformation rule for the gauge field $\ho\indices{^{i}_{k\mu}}$ is derived
by inserting the gauge Hamiltonians~(\ref{g-ham1}) and~(\ref{g-ham2}) into Eq.~(\ref{eq:inv-cond}).
Beforehand, the gauge Hamiltonian~(\ref{g-ham2}) is expressed in terms of the ``original'' (i.e.\ untransformed, lower case) fields
according to the canonical transformation rules~(\ref{F3derivative213}):
\begin{equation*}
\tilde{\HCd}_{\mathrm{Gau}_1}^{\prime}=-\tilde{k}\indices{_i^{[\mu\nu]}}
\Lambda\indices{^i_I}\,\HO\indices{^{I}_{J\alpha}}\,\Lambda\indices{^J_j}
\pfrac{X^\alpha}{x^\nu}\,e\indices{^j_\mu}\detpartial{x}{X}.
\end{equation*}
It follows that the gauge field $\ho\indices{^i_{j\nu}}$ transforms inhomogeneously as:
\begin{equation}\label{eq:omegatransform1}
\ho\indices{^i_{j\nu}}=\Lambda\indices{^i_I}\,\HO\indices{^I_{J\alpha}}\,\Lambda\indices{^J_j}\,\pfrac{X^\alpha}{x^\nu}
+\Lambda\indices{^i_I}\,\pfrac{\Lambda\indices{^I_j}}{x^{\nu}}.
\end{equation}
This transformation rule coincides with the transformation rule of spin connection coefficients---the gauge field can thus be identified with the spin connection.
With the gauge Hamiltonian~(\ref{g-ham1}), the now form-invariant action integral writes:
\begin{equation}\label{action-integral2}
S_0=\int_{V}\left[\tilde{\pi}^\nu\pfrac{\varphi}{x^\nu}+
\tilde{k}\indices{_i^{[\mu\nu]}}\left(\pfrac{e\indices{^i_\mu}}{x^\nu}+\ho\indices{^i_{j\nu}}\,e\indices{^j_\mu}\right)
-\tilde{\HCd}\right]\dd^4x.
\end{equation}
The gauge field $\ho\indices{^i_{j\nu}}(x)$ herein enters as an
external field whose dynamics is not described by the action~(\ref{action-integral2}).
This changes if we include its transformation rule~(\ref{eq:omegatransform1}) into the gauge transformation formalism.
Equation~(\ref{action-integral2}) fulfills the requirement of form-invariance under diffeomorphisms.
It also shows that no direct coupling of the scalar field $\varphi$ with the tetrad field
$e\indices{^j_\nu}$ and the gauge field $\ho\indices{^i_{j\nu}}$ emerges.
Rather, the respective coupling occurs merely via the common dependence of $\tilde{\HCd}$
and $\tilde{\HCd}_{\mathrm{Gau}_1}$ on the tetrad field $e\indices{^j_\nu}$.
The reason is that $\tilde{\pi}^\nu\,\partial\varphi/\partial x^\nu$ in the action functional~(\ref{action-integral2})
inherently constitutes a world scalar density and is, therefore, already form invariant under diffeomorphisms.
This changes as well if we include spinor fields into the canonical transformation formalism.
\subsection{Including spinors and the canonical transformation of the gauge field \texorpdfstring{$\ho\indices{^i_{j\mu}}$}{[omega]}\label{sec:complete-cantra}}
In the second step the newly introduced gauge field $\ho\indices{^i_{j\mu}}$,
defined in Eq.~(\ref{g-ham1}), will be treated as an internal quantity.
The action functional~(\ref{action-integral2}) must then be extended to also include the pertaining
momentum field, i.e.\ the tensor density $\hoc\indices{_i^{j\mu\nu}}$ conjugate to the gauge field $\ho\indices{^i_{j\mu}}$.
Moreover, we simultaneously introduce spinor fields as the source of gravitation.
Taking this into account, and substituting the scalar field by a complex spinor
field $\psi$ with the given Hamiltonian $\tilde{\HCd}_{\mathrm{D}}$, extends the action integral to:
\begin{align}
S_0=\int_{V}\Bigg(&\tilde{\kappabar}^\nu\pfrac{\psi}{x^\nu}+\pfrac{\psibar}{x^\nu}\tilde{\kappa}^\nu
+\tilde{k}\indices{_i^{\mu\nu}}\pfrac{e\indices{^i_\mu}}{x^\nu}
+\hoc\indices{_i^{\,j\mu\nu}}\pfrac{\ho\indices{^i_{j\mu}}}{x^\nu}\nonumber\\
&-\tilde{\HCd}_{\mathrm{Gau}_2}-\tilde{\HCd}_{\mathrm{D}}\Bigg)\,\dd^4x.
\label{action-integral3}
\end{align}
The task is now to determine the gauge Hamiltonian $\tilde{\HCd}_{\mathrm{Gau}_2}$
that renders the action~(\ref{action-integral3}) diffeomorphism-invariant.
In other words, $\tilde{\HCd}_{\mathrm{Gau}_2}$ is supposed to make the integrand of~(\ref{action-integral3}) into a world scalar density.
The generating function~(\ref{def:F3_Lorentzandchart}) must then be extended to define in addition the spinor and the gauge field transformation
from Eq.~(\ref{eq:omegatransform1}):
\begin{align}
\tilde{\FCd}_{3}^\nu&\big(\Psi,\tilde{\kappabar},\Psibar,\tilde{\kappa},E,\tilde{k},\HO,\hoc,x\big)\nonumber\\
&=-\tilde{\kappabar}\indices{^\nu}\,S^{-1}\,\Psi-\Psibar\,S\,\tilde{\kappa}^{\nu}-
\tilde{k}\indices{_i^{\mu\nu}}\Lambda\indices{^i_I}\,E\indices{^{I}_{\alpha}}\,\pfrac{X^\alpha}{x^\mu}\nonumber\\
&\quad-\hoc\indices{_i^{j\mu\nu}}\,\Bigg(\Lambda\indices{^i_I}\,\HO\indices{^I_{J\alpha}}\,
\Lambda\indices{^J_j}\,\pfrac{X^\alpha}{x^\mu}+\Lambda\indices{^i_I}\,\pfrac{\Lambda\indices{^I_j}}{x^\mu}\Bigg).
\label{eq:genfu-total}
\end{align}
Here $S$ and $S^{-1}$ are the spin-$\nicefrac{1}{2}$ representations of the Lorentz transformation matrix and its inverse to be specified below.
The complete set of specific rules for the generating function~(\ref{eq:genfu-total}) are:
\begin{subequations}\label{eq:all-rules}
\begin{alignat}{3}
\delta_\beta^\nu\,\psibar&\equiv-\pfrac{\tilde{\FCd}_{3}^\nu}{\tilde{\kappa}^{\beta}}
&=&\,\delta_\beta^\nu \,\Psibar\, S\\\
\tilde{\KCd}^{\nu}&\equiv-\pfrac{\tilde{\FCd}_{3}^\lambda}{\Psibar}\, \pfrac{X^\nu}{x^\lambda}\detpartial{x}{X}
&=&\,S\,\tilde{\kappa}^{\lambda}\,\pfrac{X^\nu}{x^\lambda}\detpartial{x}{X}\\
\delta_\beta^\nu\,\psi&\equiv-\pfrac{\tilde{\FCd}_{3}^\nu}{\tilde{\kappabar}\indices{^\beta}}
&=&\,\delta_\beta^\nu\,S^{-1}\,\Psi\\
\tilde{\KCdbar}\indices{^\nu}&\equiv-\pfrac{\tilde{\FCd}_{3}^\lambda}{\Psi}\, \pfrac{X^\nu}{x^\lambda}\detpartial{x}{X}
&=&\,\tilde{\kappabar}\indices{^\lambda}\,S^{-1}\,\pfrac{X^\nu}{x^\lambda}\detpartial{x}{X}\\
\delta_\beta^\nu \, e\indices{^{i}_{\mu}}&\equiv-\pfrac{\tilde{\FCd}_{3}^\nu}{\tilde{k}\indices{_i^{\mu\beta}}}
&=&\,\delta_\beta^\nu \,\Lambda\indices{^i_I}\,
E\indices{^{I}_{\alpha}}\pfrac{X^\alpha}{x^\mu}\label{eq:tet-trans}\\
\tilde{K}\indices{_I^{\mu\nu}}&\equiv-\pfrac{\tilde{\FCd}_{3}^\lambda}{E\indices{^{I}_{\mu}}}\,\pfrac{X^\nu}{x^\lambda}\detpartial{x}{X}
&=&\,\Lambda\indices{_I^i}\,\tilde{k}\indices{_i^{\xi\lambda}}\,\pfrac{X^\mu}{x^\xi}\pfrac{X^\nu}{x^\lambda}\detpartial{x}{X}
\end{alignat}
and
\begin{align}
\delta^\nu_\beta\,\ho\indices{^i_{j\mu}} &\equiv -\pfrac{\tilde{\FCd}_{3}^\nu}{\hoc\indices{_i^{j\mu\beta}}}
=\delta^\nu_\beta\Bigg(\Lambda\indices{^i_I}\,\HO\indices{^I_{J\alpha}}\,\Lambda\indices{^J_j}\pfrac{X^\alpha}{x^\mu}+
\Lambda\indices{^i_I}\,\pfrac{\Lambda\indices{^I_j}\,}{x^\mu}\Bigg)\label{eq:conn-trans}\\
\HOc\indices{_I^{J\mu\nu}} &\equiv -\pfrac{\tilde{\FCd}_{3}^\lambda}{\HO\indices{^I_{J\mu}}}
\pfrac{X^\nu}{x^\lambda} \detpartial{x}{X}
=\Lambda\indices{_I^i}\hoc\indices{_i^{j\xi\lambda}}\Lambda\indices{_j^J}\pfrac{X^\mu}{x^\xi}\pfrac{X^\nu}{x^\lambda}
\detpartial{x}{X}.\label{rhorhoLT}
\end{align}
\end{subequations}
Rule~(\ref{eq:conn-trans}) indeed reproduces the inhomogeneous transformation property of the gauge field
$\ho\indices{^i_{j\mu}}$ as required by Eq.~(\ref{eq:omegatransform1}).
The rule~(\ref{rhorhoLT}) determines the transformation property of the pertaining conjugate momentum field $\hoc\indices{_i^{j\mu\nu}}$.
\subsection{Spinor representation of the Lorentz transformation}
The parameters of the transformation given by the spinor transformation matrix $S$ are not independent
of those of the Lorentz transformation $\Lambda\indices{^I_j}$ with coefficients $\epsilon_{iJ}=-\epsilon_{Ji}$.
Rather, with the Dirac matrices $\Gamma_J$ and $\gamma_i$ in the inertial frame,
we set up the spinor representation of the Lorentz transformation as
\begin{equation}\label{eq:gamma-S}
\Gamma_J=\Lambda\indices{_J^i}\,S\,\gamma_i\,S^{-1}\qquad\Rightarrow\qquad
E\indices{_\alpha^J}\Gamma_J=\pfrac{x^\beta}{X^\alpha}\,e\indices{_\beta^i}\,S\,\gamma_i\,S^{-1}.
\end{equation}
For the commutators of the fundamental spinors,
\begin{equation}\label{eq:sigma-def}
\sigma^{ij}\equiv\ihalf\left(\gamma^i\,\gamma^j-\gamma^j\,\gamma^i\right),\qquad
\Sigma^{IJ}\equiv\ihalf\left(\Gamma^I\,\Gamma^J-\Gamma^J\,\Gamma^I\right),
\end{equation}
Eq.~(\ref{eq:gamma-S}) induces the transformation rule:
\begin{equation}
\Sigma\indices{_I^J}=\Lambda\indices{_I^i}\,S\,\sigma\indices{_i^j}\,S^{-1}\,\Lambda\indices{_j^J}.
\end{equation}
The infinitesimal representations of the local Lorentz transformation matrix $\Lambda\indices{^I_j}(x)$
and the corresponding spinor transformation matrix $S(x)$ are computed as~(see, for instance,~\cite{peskin95}):
\begin{equation}
\Lambda\indices{^I_j}=\delta_j^I+\frac{1}{2}\left(\epsilon\indices{^I_j}-\epsilon\indices{_j^I}\right),\qquad S=\Eins-\iquarter\epsilon\indices{^I_j}\sigma\indices{_I^j},
\end{equation}
where $\epsilon\indices{^I_j}(x)$ denotes the coefficients of the local Lorentz transformation.
It follows to first order in $\epsilon\indices{^I_j}$
\begin{align*}
\Lambda\indices{^i_I}\,\pfrac{\Lambda\indices{^I_j}\,}{x^\mu}&=\delta_I^i\frac{1}{2}\Bigg(\pfrac{\epsilon\indices{^I_j}}{x^\mu}-\pfrac{\epsilon\indices{_j^I}}{x^\mu}\Bigg)
=\frac{1}{2}\Bigg(\pfrac{\epsilon\indices{^i_j}}{x^\mu}-\pfrac{\epsilon\indices{_j^i}}{x^\mu}\Bigg)\\
S^{-1}\pfrac{S}{x^\mu}&=-\Eins\iquarter\pfrac{\epsilon\indices{^I_j}}{x^\mu}\sigma\indices{_I^j}
=-\iquarter\pfrac{\epsilon\indices{_i_j}}{x^\mu}\sigma\indices{^i^j}\\
&=-\frac{\rmi}{8}\,\Bigg(\pfrac{\epsilon\indices{^i_j}}{x^\mu}-\pfrac{\epsilon\indices{^i_j}}{x^\mu}\Bigg)\,\sigma\indices{_i^j},
\end{align*}
from which we conclude that
\begin{equation}\label{eq:conn-trans-spinor0}
\Lambda\indices{^i_I}\,\pfrac{\Lambda\indices{^I_j}\,}{x^\mu}\sigma\indices{_i^j}=-\frac{4}{\rmi}\,S^{-1}\pfrac{S}{x^\mu}.
\end{equation}
This yields the spinor representation of the transformation rule~(\ref{eq:omegatransform1}) for the gauge field $\ho\indices{^i_{j\mu}}$,
\begin{align*}
S\ho\indices{^i_{j\mu}}\sigma\indices{_i^j}S^{-1}\!&=\Lambda\indices{^i_I}S\HO\indices{^I_{J\nu}}\Lambda\indices{^J_j}\sigma\indices{_i^j}S^{-1}\pfrac{X^\nu}{x^\mu}
+S\Lambda\indices{^i_I}\pfrac{\Lambda\indices{^I_j}}{x^{\mu}}\sigma\indices{_i^j}S^{-1}\\
&=\HO\indices{^I^J_\nu}\,\Lambda\indices{_i_I}\,S\,\sigma\indices{^i^j}\,S^{-1}\,\Lambda\indices{_J_j}\,\pfrac{X^\nu}{x^\mu}-\frac{4}{\rmi}\pfrac{S}{x^\mu}S^{-1}\\
&=\HO\indices{^I^J_\nu}\,\Lambda\indices{_I_i}\,S\,\sigma\indices{^i^j}\,S^{-1}\,\Lambda\indices{_j_J}\,\pfrac{X^\nu}{x^\mu}-\frac{4}{\rmi}\pfrac{S}{x^\mu}S^{-1}\\
&=\HO\indices{^I_J_\nu}\,\Sigma\indices{_I^J}\,\pfrac{X^\nu}{x^\mu}-\frac{4}{\rmi}\pfrac{S}{x^\mu}S^{-1},
\end{align*}
hence, in analogy to Eq.~(\ref{eq:conn-trans}):
\begin{equation}\label{eq:conn-trans-spinor}
\iquarter\HO\indices{^I_{J\nu}}\Sigma\indices{_I^J}=\left(\iquarter\,S\,\ho\indices{^i_{j\mu}}\,\sigma\indices{_i^j}\,S^{-1}+\pfrac{S}{x^\mu}S^{-1}\right)\pfrac{x^\mu}{X^\nu}.
\end{equation}
\subsection{Derivation of the gauge Hamiltonian}
The key benefit of the canonical transformation framework is that it provides
the prescription for gauging the initial Hamiltonian density $\tilde{\HCd}_{\mathrm{D}}$,
hence to derive the gauge Hamiltonian $\tilde{\HCd}_{\mathrm{Gau}}$ such that the
combined system $\tilde{\HCd}_{\mathrm{D}}+\tilde{\HCd}_{\mathrm{Gau}}$ becomes diffeomorphism-invariant.
The gauge Hamiltonian is ultimately determined by the explicit
$x^\mu$-dependence of the generating function according to the general rule~(\ref{eq:f3derivative}).
For the actual generating function~(\ref{eq:genfu-total}), the $x^\nu$-derivative
of the space-time dependent parameters in the generating function evaluates to:
\begin{align}
\pfrac{\tilde{\FCd}_{3}^\nu}{x^\nu}\bigg|_{\text{expl}}\!\!\!\!\!&=
-\tilde{\kappabar}\indices{^\nu}\,\pfrac{S^{-1}}{x^\nu}\,\Psi-\Psibar\,\pfrac{S}{x^\nu}\,\tilde{\kappa}^{\nu}
-\tilde{k}\indices{_i^{\mu\nu}}\pfrac{}{x^\nu}\left(\Lambda\indices{^i_I}\,\pfrac{X^\alpha}{x^\mu}\right)E\indices{^{I}_{\alpha}}\nonumber\\
&-\hoc\indices{^{[ij]\mu\nu}}\!\Bigg[\HO\indices{^I_{J\alpha}}\pfrac{}{x^\nu}
\Bigg(\!\Lambda\indices{_i_I}\Lambda\indices{^J_j}\pfrac{X^\alpha}{x^\mu}\!\Bigg)
+\pfrac{}{x^\nu}\Bigg(\!\Lambda\indices{_i_I}\pfrac{\Lambda\indices{^I_j}}{x^\mu}\!\Bigg)\Bigg].
\label{eq:f3derivative2}
\end{align}
The final gauge Hamiltonian is then obtained from~(\ref{eq:f3derivative2}) by expressing all its parameters, namely
$\Lambda\indices{^i_I}$, $S$, $\partial X^\alpha/\partial x^\mu$ and their respective derivatives, in terms
of the physical fields of the system according to the set of canonical transformation rules~(\ref{eq:all-rules}).
This will be worked out in the following subsections.
\subsubsection{Contribution of the spinor fields \texorpdfstring{$\Psi,\Psibar$}{[Psi]} to Eq.~(\ref{eq:f3derivative2})}
By means of the transformation rule~(\ref{eq:conn-trans-spinor}) for $S$, the replacement
of the derivative in the first (spinor) term of~Eq.~(\ref{eq:f3derivative2}) follows as:
\begin{align*}
-\tilde{\kappabar}^\nu\pfrac{S^{-1}}{x^\nu}\,\Psi
&=\tilde{\kappabar}^\nu S^{-1}\,\pfrac{S}{x^\nu}\,\psi\\
&=\iquarter\tilde{\kappabar}^\nu\Big(S^{-1}\,\HO\indices{^I_{J\alpha}}\,
\Sigma\indices{_I^J}\,S\,\pfrac{X^\alpha}{x^\nu}-\ho\indices{^i_{j\nu}}\,\sigma\indices{_i^j}\Big)\,\psi\\
&=\tilde{\KCdbar}^\nu\,\iquarter\HO\indices{^I_{J\nu}}\,\Sigma\indices{_I^J}\,\Psi\detpartial{X}{x}
-\tilde{\kappabar}^\nu\,\iquarter\ho\indices{^i_{j\nu}}\,\sigma\indices{_i^j}\,\psi.
\end{align*}
Similarly for the second term:
\begin{align*}
-\Psibar\,\pfrac{S}{x^\nu}\,\tilde{\kappa}^{\nu}
&=-\psibar\,S^{-1}\,\pfrac{S}{x^\nu}\,\tilde{\kappa}^{\nu}\\
&=\psibar\,\iquarter\ho\indices{^i_{j\nu}}\,\sigma\indices{_i^j}\,\tilde{\kappa}^{\nu}
-\Psibar\,\iquarter\HO\indices{^I_{J\nu}}\,\Sigma\indices{_I^J}\tilde{\KCd}^{\nu}\detpartial{X}{x}.
\end{align*}
Hence, the free parameters of the spinor-related terms in Eq.~(\ref{eq:f3derivative2})
are replaced by the connection fields according to
\begin{align*}
-\tilde{\kappabar}\indices{^\nu}\,\pfrac{S^{-1}}{x^\nu}\,\Psi&-\Psibar\,\pfrac{S}{x^\nu}\,\tilde{\kappa}^{\nu}
=\iquarter\left(\psibar\,\ho\indices{^i_{j\nu}}\,\sigma\indices{_i^j}\,\tilde{\kappa}^{\nu}-
\tilde{\kappabar}\indices{^\nu}\,\ho\indices{^i_{j\nu}}\,\sigma\indices{_i^j}\,\psi\right)\\
&-\iquarter\Big(\Psibar\,\HO\indices{^I_{J\nu}}\,\Sigma\indices{_I^J}\,\tilde{\KCd}^{\nu}-
\tilde{\KCdbar}\indices{^\nu}\,\HO\indices{^I_{J\nu}}\,\Sigma\indices{_I^J}\,\Psi\Big)\detpartial{X}{x}.
\end{align*}
\subsubsection{Contribution of the tetrad field \texorpdfstring{$E\indices{^I_\alpha}$}{E} to Eq.~(\ref{eq:f3derivative2})}
The parameters in the (third) term proportional to $\tilde{k}\indices{_i^{\mu\nu}}$ can, with help of the transformation
rules~(\ref{eq:all-rules}), be similarly expressed in terms of the physical fields as follows:
\begin{align*}
&\quad\,\tilde{k}\indices{_i^{\mu\nu}}\,\pfrac{}{x^\nu}\left(\Lambda\indices{^i_J}
\pfrac{X^\alpha}{x^\mu}\right)E\indices{^{J}_{\alpha}}\nonumber\\
&=\onehalf\tilde{k}\indices{_i^{\mu\nu}}\left(\pfrac{e\indices{^{i}_{\mu}}}{x^\nu}+\pfrac{e\indices{^{i}_{\nu}}}{x^\mu}-
\ho\indices{^i_{j\nu}}e\indices{^{j}_{\mu}}+\ho\indices{^i_{j\mu}}e\indices{^{j}_{\nu}}\right)\nonumber\\
&\quad-\onehalf\tilde{K}\indices{_I^{\mu\nu}}\,\Bigg(\pfrac{E\indices{^{I}_{\mu}}}{X^\nu}+\pfrac{E\indices{^{I}_{\nu}}}{X^\mu}-
\HO\indices{^I_{J\nu}}E\indices{^{J}_{\mu}}+\HO\indices{^I_{J\mu}}E\indices{^{J}_{\nu}}\Bigg)\detpartial{X}{x}.
\end{align*}
The detailed calculation of this result is worked out in App.~\ref{app:tetrad-calc}.
\subsubsection{Contribution of the gauge field \texorpdfstring{$\HO\indices{^I_{J\alpha}}$}{[Omega]} to Eq.~(\ref{eq:f3derivative2})}
As the last step, we express the coefficients in the term proportional to $\hoc\indices{_i^{j\mu\nu}}$ of Eq.~(\ref{eq:f3derivative2})
in terms of the physical fields according to the canonical transformation rules~(\ref{eq:all-rules}):
\begin{align}
&\quad-\hoc\indices{_i^{j\mu\nu}}\Bigg[\HO\indices{^I_{J\alpha}}\,\pfrac{}{x^\nu}
\left(\Lambda\indices{^i_I}\,\Lambda\indices{^J_j}\,\pfrac{X^\alpha}{x^\mu}\right)
+ \pfrac{}{x^\nu}\,\Bigg(\Lambda\indices{^i_I}\,\pfrac{\Lambda\indices{^I_j}}{x^\mu}\Bigg)\Bigg]\nonumber\\
&=-\onehalf\hoc\indices{_i^{j\mu\nu}}\Bigg(
\pfrac{\ho\indices{^i_{j\mu}}}{x^\nu}+\pfrac{\ho\indices{^i_{j\nu}}}{x^\mu}
+\ho\indices{^i_{n\mu}}\,\ho\indices{^n_{j\nu}}-\ho\indices{^i_{n\nu}}\,\ho\indices{^n_{j\mu}}\Bigg)\nonumber\\
&+\onehalf\HOc\indices{_I^{J\mu\nu}}\!\left(\pfrac{\HO\indices{^I_{J\mu}}}{X^\nu}+\pfrac{\HO\indices{^I_{J\nu}}}{X^\mu}
+\HO\indices{^I_{N\mu}}\HO\indices{^N_{J\nu}}\!-\HO\indices{^I_{N\nu}}\HO\indices{^N_{J\mu}}\!\right)\detpartial{X}{x}.
\label{eq:f3derivative3}
\end{align}
The detailed calculation is worked out in App.~\ref{app:gauge-calc}.
\allowdisplaybreaks\relax
\subsubsection{Gauge Hamiltonian}
The replacement of the explicitly $x^\mu$-dependent parameters in the divergence~(\ref{eq:f3derivative2})
of the generating function by the actual physical fields now sums up to
\begin{align}
\pfrac{\tilde{\FCd}_{3}^\nu}{x^\nu}&\bigg|_{\text{expl}}
=-\iquarter\left(\tilde{\kappabar}^\nu\,\ho\indices{^i_{j\nu}}\,\sigma\indices{_i^j}\,\psi
-\psibar\,\ho\indices{^i_{j\nu}}\,\sigma\indices{_i^j}\,\tilde{\kappa}^\nu\right)\nonumber\\
&-\onehalf\tilde{k}\indices{_i^{\mu\nu}}\left(\pfrac{e\indices{^i_\mu}}{x^\nu}+\pfrac{e\indices{^i_\nu}}{x^\mu}
+\ho\indices{^i_{j\mu}}\,e\indices{^j_\nu}-\ho\indices{^i_{j\nu}}\,e\indices{^j_\mu}\right)\nonumber\\
&-\onehalf\hoc\indices{_i^{j\mu\nu}}\Bigg(
\pfrac{\ho\indices{^i_{j\mu}}}{x^\nu}+\pfrac{\ho\indices{^i_{j\nu}}}{x^\mu}
+\ho\indices{^i_{n\mu}}\,\ho\indices{^n_{j\nu}}-\ho\indices{^i_{n\nu}}\,\ho\indices{^n_{j\mu}}\Bigg)\nonumber\\
&+\iquarter\left(\tilde{\KCdbar}\indices{^\nu}\,\HO\indices{^I_{J\nu}}\,\Sigma\indices{_I^J}\,\Psi
-\Psibar\,\HO\indices{^I_{J\nu}}\,\Sigma\indices{_I^J}\tilde{\KCd}^{\nu}\right)\nonumber\\
&+\onehalf\tilde{K}\indices{_I^{\mu\nu}}\left(\pfrac{E\indices{^{I}_{\mu}}}{X^\nu}+\pfrac{E\indices{^{I}_{\nu}}}{X^\mu}
+\HO\indices{^I_{J\mu}}\,E\indices{^J_\nu}-\HO\indices{^I_{J\nu}}\,E\indices{^J_\mu}\right)\nonumber\\
&+\onehalf\HOc\indices{_I^{J\mu\nu}}\!\left(\pfrac{\HO\indices{^I_{J\mu}}}{X^\nu}+\pfrac{\HO\indices{^I_{J\nu}}}{X^\mu}
+\HO\indices{^I_{N\mu}}\HO\indices{^N_{J\nu}}\!-\HO\indices{^I_{N\nu}}\HO\indices{^N_{J\mu}}\!\right).
\label{eq:f3-div-total}
\end{align}
The partial derivatives associated with $\tilde{k}\indices{_i^{\mu\nu}}$ and
$\hoc\indices{_i^{j\mu\nu}}$  in~(\ref{eq:f3-div-total}) are merged with
the corresponding derivatives contained in the initial action functional~(\ref{action-integral3})
to yield the following modified action functional
\begin{align}
S_0=\int_{V}&\!\dd^4x\,\Bigg[\tilde{\kappabar}^\nu\pfrac{\psi}{x^{\nu}}+\pfrac{\psibar}{x^{\nu}}\tilde{\kappa}^\nu
+\onehalf\tilde{k}\indices{_i^{\mu\nu}}\left(\pfrac{e\indices{^{i}_{\mu}}}{x^{\nu}}-\pfrac{e\indices{^{i}_{\nu}}}{x^{\mu}}\right)\nonumber\\
&\!+\onehalf\hoc\indices{_i^{\,j\mu\nu}}\Bigg(\pfrac{\ho\indices{^i_{j\mu}}}{x^{\nu}}-\pfrac{\ho\indices{^i_{j\nu}}}{x^{\mu}}\Bigg)
-\tilde{\HCd}_{\mathrm{Gau}_2}\!-\tilde{\HCd}_{\mathrm{D}}\Bigg].
\label{action-integral4}
\end{align}
Then the total gauge Hamiltonian $\tilde{\HCd}_{\mathrm{Gau}_2}$---which is form-invariant under the transformation
rule~(\ref{eq:f3derivative}) for the transformation of spinor fields in a curvilinear space-time
defined by the generating function~(\ref{eq:f3-div-total})---follows as
\begin{align}
&\quad\tilde{\HCd}_{\mathrm{Gau}_2}
=\iquarter\tilde{\kappabar}\indices{^\beta}\,\ho\indices{^i_{j\beta}}\,\sigma\indices{_i^j}\,\psi
-\iquarter\psibar\,\ho\indices{^i_{j\beta}}\,\sigma\indices{_i^j}\,\tilde{\kappa}^{\beta}\nonumber\\
&+\onehalf\tilde{k}\indices{_i^{\alpha\beta}}\!\left(
\ho\indices{^{i}_{j\alpha}}e\indices{^{j}_{\beta}}\!
-\ho\indices{^{i}_{j\beta}}e\indices{^{j}_{\alpha}}\right)
+\!\onehalf\hoc\indices{_{i}^{j\alpha\beta}}\!\left(
\ho\indices{^{i}_{n\alpha}}\ho\indices{^{n}_{j\beta}}\!
-\ho\indices{^{i}_{n\beta}}\ho\indices{^{n}_{j\alpha}}\right).
\label{eq:final-gauge-ham}
\end{align}
\subsubsection{Generally covariant action in dynamical space-time\label{sec:ge-cov-act}}
Inserting the gauge Hamiltonian~(\ref{eq:final-gauge-ham}) into the action integral~(\ref{action-integral4})
yields the final form-invariant action functional.
It involves the Hamiltonian
$\tilde{\HCd}_{\mathrm{D}}\big(\tilde{\kappa},\psibar,\tilde{\kappabar},\psi,e\big)$ from Eq.~(\ref{hd-dirac})
of the free (uncoupled) system of complex spinor fields,
and the Hamiltonian  $\tilde{\HCd}_{\mathrm{Gr}}\big(\tilde{k},e,\hoc\,\big)$
of the free gravitational field:
\begin{align}
S_0=\!\int_{V}&\Bigg[\,
\tilde{\kappabar}^\nu\left(\pfrac{\psi}{x^\nu}-\iquarter\ho\indices{^i_{j\nu}}\sigma\indices{_i^j}\psi\right)
+\left(\pfrac{\psibar}{x^\nu}+\iquarter\ho\indices{^i_{j\nu}}\psibar\sigma\indices{_i^j}\right)\tilde{\kappa}^\nu\nonumber\\
&+\frac{1}{2}\tilde{k}\indices{_i^{\mu\nu}}\left(\pfrac{e\indices{^i_\mu}}{x^\nu}-\pfrac{e\indices{^i_\nu}}{x^\mu}
+\ho\indices{^i_{j\nu}}\,e\indices{^j_\mu}-\ho\indices{^i_{j\mu}}\,e\indices{^j_\nu}\right)\nonumber\\
&+\frac{1}{2}\hoc\indices{_i^{j\mu\nu}}\Bigg(\pfrac{\ho\indices{^i_{j\mu}}}{x^\nu}
-\pfrac{\ho\indices{^i_{j\nu}}}{x^\mu}+\ho\indices{^i_{n\nu}}\,\ho\indices{^n_{j\mu}}
-\ho\indices{^i_{n\mu}}\,\ho\indices{^n_{j\nu}}\Bigg)\nonumber\\
&-\tilde{\HCd}_{\mathrm{D}}-\tilde{\HCd}_{\mathrm{Gr}}\Bigg]\,\dd^4x.
\label{action-integral5}
\end{align}
The particular choice of $\tilde{\HCd}_{\mathrm{Gr}}\big(\tilde{k},e,\hoc\,\big)$ must be based on physical reasoning.
It determines the version of the space-time,
i.e.\ flat, Riemann, or Riemann-Cartan and its dynamics in the absence of any external sources.
$S_0$ does not contain external functions anymore and thus represents a closed physical system of fermions in a dynamical space-time.
\section{Canonical field equations}
\subsection{Canonical equations for the spinor field}
The set of canonical field equations for the spinors $\psi$ and $\psibar$ follows from the action~(\ref{action-integral5}) as:
\begin{subequations}\label{eq:spinor-feqs}
\begin{align}
\pfrac{\psi}{x^\nu}&=\hphantom{-}\pfrac{\tilde{\HCd}_{\mathrm{D}}}{\tilde{\kappabar}^\nu}+\iquarter\,\ho\indices{^i_{j\nu}}\,\sigma\indices{_i^j}\,\psi\\
\pfrac{\tilde{\kappabar}^\nu}{x^\nu}&=-\pfrac{\tilde{\HCd}_{\mathrm{D}}}{\psi}-\iquarter\,\ho\indices{^i_{j\nu}}\,\tilde{\kappabar}^\nu\,\sigma\indices{_i^j}\\
\pfrac{\psibar}{x^\nu}&=\hphantom{-}\pfrac{\tilde{\HCd}_{\mathrm{D}}}{\tilde{\kappa}^\nu}-\iquarter\,\ho\indices{^i_{j\nu}}\,\psibar\,\sigma\indices{_i^j}\\
\pfrac{\tilde{\kappa}^\nu}{x^\nu}&=-\pfrac{\tilde{\HCd}_{\mathrm{D}}}{\psibar}+\iquarter\,\ho\indices{^i_{j\nu}}\,\sigma\indices{_i^j}\,\tilde{\kappa}^\nu.
\end{align}
\end{subequations}
\subsection{Canonical equations for the tetrad field}
The canonical equation for the derivative of the tetrad $e\indices{^i_{\mu}}$ follows as
\begin{equation*}
\pfrac{e\indices{^i_\mu}}{x^\nu}-\pfrac{e\indices{^i_\nu}}{x^\mu}=
2\pfrac{\tilde{\HCd}_\mathrm{Gr}}{\tilde{k}\indices{_i^{\mu\nu}}}+
\ho\indices{^i_{j\mu}}\,e\indices{^j_\nu}-\ho\indices{^i_{j\nu}}\,e\indices{^j_\mu},
\end{equation*}
and thus
\begin{align}
\pfrac{\tilde{\HCd}_\mathrm{Gr}}{\tilde{k}\indices{_i^{\mu\nu}}}&=\onehalf\left(\pfrac{e\indices{^i_{\mu}}}{x^\nu}
-\pfrac{e\indices{^i_\nu}}{x^\mu}+\ho\indices{^i_{j\nu}}\,e\indices{^j_\mu}-\ho\indices{^i_{j\mu}}\,e\indices{^j_\nu}\right).
\label{eq:e-deri}
\end{align}
The conjugate canonical equation~(\ref{eq:feq-KG-4}) for the divergence of $\tilde{k}\indices{_i^{\mu\nu}}$ is then:
\begin{align*}
\pfrac{\tilde{k}\indices{_i^{[\mu\alpha]}}}{x^\alpha}&=
-\pfrac{\tilde{\HCd}_{\mathrm{D}}}{e\indices{^i_{\mu}}}-\pfrac{\tilde{\HCd}_\mathrm{Gr}}{e\indices{^i_{\mu}}}
-\pfrac{\tilde{\HCd}_\mathrm{Gau_2}}{e\indices{^i_{\mu}}}\\
&=-\pfrac{\tilde{\HCd}_{\mathrm{D}}}{e\indices{^i_{\mu}}}-\pfrac{\tilde{\HCd}_{\mathrm{Gr}}}{e\indices{^i_{\mu}}}
+\onehalf\left(\tilde{k}\indices{_j^{\mu\nu}}-\tilde{k}\indices{_j^{\nu\mu}}\right)\ho\indices{^{j}_{i\nu}}\\
&=-\pfrac{\tilde{\HCd}_{\mathrm{D}}}{e\indices{^i_{\mu}}}-\pfrac{\tilde{\HCd}_{\mathrm{Gr}}}{e\indices{^i_{\mu}}}
+\tilde{k}\indices{_j^{[\mu\alpha]}}\,\ho\indices{^{j}_{i\alpha}}.
\end{align*}
Regrouping the terms yields:
\begin{equation}\label{eq:k-div3}
\Bigg(\pfrac{\tilde{k}\indices{_i^{[\mu\alpha]}}}{x^\alpha}-\tilde{k}\indices{_j^{[\mu\alpha]}}\,\ho\indices{^{j}_{i\alpha}}\Bigg)\,e\indices{^i_{\nu}}
=-\pfrac{\tilde{\HCd}_{\mathrm{D}}}{e\indices{^i_{\mu}}}e\indices{^i_{\nu}}
-\pfrac{\tilde{\HCd}_{\mathrm{Gr}}}{e\indices{^i_{\mu}}}e\indices{^i_{\nu}}.
\end{equation}
The right-hand side of Eq.~(\ref{eq:k-div3}) is exactly the Hamiltonian representation of the metric
energy-momentum tensors of the source system described by $\tilde{\HCd}_{\mathrm{D}}$,
and of the free gravitational field specified by a Hamiltonian $\tilde{\HCd}_{\mathrm{Gr}}$.
\subsection{Canonical equations for the connection field}
Taking into account the skew-symmetry $\ho\indices{_i_{j\mu}}=-\ho\indices{_j_{i\mu}}$ of the connection,
the canonical equation for the divergence of $\hoc\indices{_i^{j\mu\nu}}$ follows as
\begin{align}
\pfrac{\hoc\indices{_i^{j\mu\nu}}}{x^\nu}&=-\pfrac{\tilde{\HCd}_{\mathrm{Gau}_2}}{\ho\indices{^i_{j\mu}}}\nonumber\\
&=-\iquarter\tilde{\kappabar}^\mu\,\sigma\indices{_i^j}\,\psi+\iquarter\psibar\,\sigma\indices{_i^j}\,\tilde{\kappa}^\mu
-\onehalf\left(\tilde{k}\indices{_i^{\mu\nu}}-\tilde{k}\indices{_i^{\nu\mu}}\right)e\indices{^j_\nu}\nonumber\\
&\quad+\onehalf\left(\hoc\indices{_i^{n\nu\mu}}-\hoc\indices{_i^{n\mu\nu}}\right)\ho\indices{^j_{n\nu}}
-\onehalf\left(\hoc\indices{_n^{j\nu\mu}}-\hoc\indices{_n^{j\mu\nu}}\right)\ho\indices{^n_{i\nu}}.
\label{eq:t-div}
\end{align}
The canonical equation for the derivative of the gauge field $\ho\indices{^i_{j\mu}}$ follows as
\begin{equation*}
\pfrac{\ho\indices{^i_{j\mu}}}{x^\nu}-\pfrac{\ho\indices{^i_{j\nu}}}{x^\mu}
=2\pfrac{\tilde{\HCd}_\mathrm{Gr}}{\hoc\indices{_i^{j\mu\nu}}}+
\ho\indices{^i_{n\mu}}\,\ho\indices{^n_{j\nu}}-\ho\indices{^i_{n\nu}}\,\ho\indices{^n_{j\mu}}.
\end{equation*}
Combining the four spin connection terms gives
the mixed representation of the Riemann-Cartan curvature tensor $R\indices{^i_{j\nu\mu}}$ (see~App.~\ref{app:gamma_connection}):
\begin{align}
\pfrac{\tilde{\HCd}_\mathrm{Gr}}{\hoc\indices{_i^{j\mu\nu}}}&=
\onehalf\Bigg(\pfrac{\ho\indices{^i_{j\mu}}}{x^\nu}-\pfrac{\ho\indices{^i_{j\nu}}}{x^\mu}
+\ho\indices{^i_{n\nu}}\,\ho\indices{^n_{j\mu}}-\ho\indices{^i_{n\mu}}\,\ho\indices{^n_{j\nu}}\Bigg)\nonumber\\
&=\onehalf R\indices{^i_{j\nu\mu}}.
\label{eq:omega-deri}
\end{align}
\subsection{Summary of the coupled set of canonical field equations}
Below we finally summarize the complete closed set of eight coupled field equations for a system of spinor fields
in a dynamical space-time resulting from the variation of the action functional~(\ref{action-integral5}):
\begin{subequations}\label{eq:feqs-all}
\begin{align}
\pfrac{\psi}{x^\nu}&=\hphantom{-}\pfrac{\tilde{\HCd}_{\mathrm{D}}}{\tilde{\kappabar}^\nu}
+\iquarter\,\ho\indices{^i_{j\nu}}\,\sigma\indices{_i^j}\,\psi\label{eq:psi-deri2}\\
\pfrac{\tilde{\kappabar}^\alpha}{x^\alpha}&=-\pfrac{\tilde{\HCd}_{\mathrm{D}}}{\psi}
-\iquarter\,\ho\indices{^i_{j\alpha}}\,\tilde{\kappabar}^\alpha\,\sigma\indices{_i^j}\label{eq:kappabar-div2}\\
\pfrac{\psibar}{x^\nu}&=\hphantom{-}\pfrac{\tilde{\HCd}_{\mathrm{D}}}{\tilde{\kappa}^\nu}
-\iquarter\,\ho\indices{^i_{j\nu}}\,\psibar\,\sigma\indices{_i^j}\label{eq:psibar-deri2}\\
\pfrac{\tilde{\kappa}^\alpha}{x^\alpha}&=-\pfrac{\tilde{\HCd}_{\mathrm{D}}}{\psibar}
+\iquarter\,\ho\indices{^i_{j\alpha}}\,\sigma\indices{_i^j}\,\tilde{\kappa}^\alpha\label{eq:kappa-div2}\\
\pfrac{\tilde{k}\indices{_i^{\mu\alpha}}}{x^\alpha}&=
-\pfrac{\tilde{\HCd}_{\mathrm{D}}}{e\indices{^i_{\mu}}}-\pfrac{\tilde{\HCd}_{\mathrm{Gr}}}{e\indices{^i_{\mu}}}
+\tilde{k}\indices{_j^{[\mu\alpha]}}\ho\indices{^{j}_{i\alpha}}\label{eq:k-div2}\\
\pfrac{e\indices{^i_\mu}}{x^\nu}-\pfrac{e\indices{^i_\nu}}{x^\mu}
&=\hphantom{-}2\pfrac{\tilde{\HCd}_\mathrm{Gr}}{\tilde{k}\indices{_i^{\mu\nu}}}+
\ho\indices{^i_{j\mu}}\,e\indices{^j_\nu}-\ho\indices{^i_{j\nu}}\,e\indices{^j_\mu}\label{eq:e-deri2}\\
\pfrac{\ho\indices{^i_{j\mu}}}{x^\nu}-\pfrac{\ho\indices{^i_{j\nu}}}{x^\mu}
&=\hphantom{-}2\pfrac{\tilde{\HCd}_\mathrm{Gr}}{\hoc\indices{_i^{j\mu\nu}}}+
\ho\indices{^i_{n\mu}}\,\ho\indices{^n_{j\nu}}-\ho\indices{^i_{n\nu}}\,\ho\indices{^n_{j\mu}}\label{eq:omega-deri2}\\
\pfrac{\hoc\indices{_i^{j\mu\alpha}}}{x^\alpha}&=-\pfrac{\tilde{\HCd}_{\mathrm{Gau}_2}}{\ho\indices{^i_{j\mu}}}
=\hoc\indices{_i^{n[\alpha\mu]}}\,\ho\indices{^j_{n\alpha}}-\hoc\indices{_n^{j[\alpha\mu]}}\,\ho\indices{^n_{i\alpha}}\nonumber\\
&\qquad+\tilde{k}\indices{_i^{[\alpha\mu]}}\,e\indices{^j_\alpha}
-\iquarter\tilde{\kappabar}^\mu\,\sigma\indices{_i^j}\,\psi
+\iquarter\psibar\,\sigma\indices{_i^j}\,\tilde{\kappa}^\mu.\label{eq:t-div-2}
\end{align}
\end{subequations}
This set of canonical equations provides a self-consistent description of the coupled dynamics of spinor fields and space-time.
It extends Eqs.~(\ref{eq:feq-KG-3}), (\ref{eq:feq-KG-4}), and~(\ref{eq:can-dirac-1-4}), which merely take into account the curvilinearity of the space-time geometry.
\subsection{Dirac equation with coupling to the connection field \texorpdfstring{$\ho\indices{^i_{j\nu}}$}{[omega]}}
Inserting the partial differential equations for the spinor fields~(\ref{eq:can-dirac-1-4})
into Eqs.~(\ref{eq:psi-deri2}) to~(\ref{eq:kappa-div2}),
gives the corresponding covariant field equations due to their coupling to the gauge field $\ho\indices{^i_{j\nu}}$
\begin{subequations}
\begin{align}
\pfrac{\psi}{x^\nu}&=-\ihalf M\left(e\indices{_\nu^i}\,\gamma_i\,\psi+e\indices{_\nu^i}\,
\frac{6\tau_{ij}}{\dete}\,e\indices{^j_\beta}\,\tilde{\kappa}^\beta\right)+\iquarter\,\ho\indices{^i_{j\nu}}\,\sigma\indices{_i^j}\,\psi\label{eq:can-dirac1b}\\
\pfrac{\tilde{\kappa}^\alpha}{x^\alpha}&=-\left(\ihalf M\gamma_i\,e\indices{^i_\alpha}
-\iquarter\,\ho\indices{^i_{j\alpha}}\,\sigma\indices{_i^j}\right)\tilde{\kappa}^\alpha-\left(m-M\right)\psi\,\dete\label{eq:can-dirac2b}\\
\pfrac{\psibar}{x^{\nu}}&=\hphantom{-}\ihalf M\left(\psibar\,\gamma_j\,e\indices{^j_\nu}-\tilde{\kappabar}^\beta\,
e\indices{_\beta^i}\,\frac{6\tau_{ij}}{\dete}\,e\indices{^j_\nu}\right)-\iquarter\,\ho\indices{^i_{j\nu}}\,\psibar\,\sigma\indices{_i^j}\label{eq:can-dirac3b}\\
\pfrac{\tilde{\kappabar}^{\alpha}}{x^{\alpha}}&=\hphantom{-}\tilde{\kappabar}^\alpha\left(\ihalf M\,\gamma_i\,e\indices{^i_\alpha}
-\iquarter\,\ho\indices{^i_{j\alpha}}\,\sigma\indices{_i^j}\right)-\left(m-M\right)\psibar\,\dete.\label{eq:can-dirac4b}
\end{align}
\end{subequations}
To express the coupled set of first order equations as a second order equation for the spinor $\psi$,
we solve Eq.~(\ref{eq:can-dirac1b}) for the spinor momentum field $\tilde{\kappa}^\alpha$,
\begin{equation}\label{eq:can-dirac1c}
\tilde{\kappa}^\alpha=e\indices{^\alpha_j}\left[-\ihalf \gamma^j\psi+\frac{\rmi}{3M}\,\sigma^{ji}\,e\indices{_i^\beta}
\left(\pfrac{\psi}{x^\beta}-\iquarter\ho\indices{^n_{m\beta}}\,\sigma\indices{_n^m}\,\psi\right)\right]\dete
\end{equation}
and insert it into Eq.~(\ref{eq:can-dirac2b}).
The explicit derivation of the generalized Dirac equation is worked out in App.~\ref{app:gen-dirac}.
The final result is:
\begin{align}
&\frac{\rmi}{3M}\sigma^{ji}\Bigg[\Bigg(\pfrac{e\indices{^\alpha_j}}{x^\alpha}e\indices{_i^\beta}
-e\indices{^\alpha_k}\ho\indices{^k_{j\alpha}}e\indices{_i^\beta}
+e\indices{^\alpha_j}\pfrac{e\indices{_i^\beta}}{x^\alpha}\nonumber\\
&\qquad\quad\!-e\indices{^\alpha_j}e\indices{^\beta_k}\ho\indices{^k_{i\alpha}}\!
-e\indices{^\alpha_j}e\indices{_i^\beta}e\indices{^k_\xi}\pfrac{e\indices{^\xi_k}}{x^\alpha}\!\Bigg)
\left(\pfrac{\psi}{x^\beta}-\iquarter\ho\indices{^n_{m\beta}}\sigma\indices{_n^m}\psi\right)\nonumber\\
&\qquad\quad-\iquarter e\indices{^\alpha_j}e\indices{_i^\beta}\sigma\indices{_n^m}
\left(\pfrac{\ho\indices{^n_{m\beta}}}{x^\alpha}+\ho\indices{^n_{k\alpha}}\ho\indices{^k_{m\beta}}\right)\psi\Bigg]\nonumber\\
&=\rmi\gamma^j e\indices{_j^\beta}\left(\pfrac{\psi}{x^\beta}-\iquarter\ho\indices{^n_{m\beta}}\sigma\indices{_n^m}\psi\right)-m\,\psi\nonumber\\
&\qquad+\ihalf\gamma^j\,\Bigg(\pfrac{e\indices{^\alpha_j}}{x^\alpha}-e\indices{^\alpha_k}\ho\indices{^k_{j\alpha}}
-e\indices{^\alpha_j}e\indices{^k_\xi}\pfrac{e\indices{^\xi_k}}{x^\alpha}\Bigg)\,\psi.
\label{eq:gen-dirac-final}
\end{align}
Due to the skew-symmetry of $\sigma^{ji}$, term proportional to $\sigma\indices{_n^m}$ is actually half
the Riemann-Cartan curvature tensor in the mixed Lorentz-coordinate space representation from Eq.~(\ref{eq:omega-deri}).
The generalized Dirac equation~(\ref{eq:gen-dirac-final}) with metric compatibility will be discussed in the following section.
\subsection{Generalized Dirac equation with metric compatibility}
It is convenient to define a new set of coefficients $\gamma\indices{^{\xi}_{\mu\nu}}$
as functions of the spin connection and tetrad fields as follows:
\begin{equation}\label{def:gammainomega2}
\gamma\indices{^{\mu}_{\alpha\nu}}\equiv
-\left(\pfrac{e\indices{^{\mu}_{i}}}{x^{\nu}}-e\indices{^{\mu}_{k}}\,\ho\indices{^{k}_{i\nu}}\right)e\indices{^{i}_{\alpha}}
=e\indices{^{\mu}_{k}}\left(\pfrac{e\indices{^{k}_{\alpha}}}{x^{\nu}}+\ho\indices{^{k}_{i\nu}}\,e\indices{^{i}_{\alpha}}\right).
\end{equation}
The transformation relation for $\gamma\indices{^{\mu}_{\alpha\nu}}$ is uniquely determined by the
transformation relation~(\ref{eq:omegatransform1}) of the skew-symmetric spin connection under arbitrary diffeomorphisms.
A straightforward calculation (c.f.~App.~\ref{app:gamma_connection}) gives:
\begin{equation}\label{gammatransforminverse}
\gamma\indices{^{\alpha}_{\nu\beta}}=\pfrac{X^\eta}{x^\nu}\pfrac{X^\mu}{x^\beta}\pfrac{x^\alpha}{X^\xi}\Gamma\indices{^{\xi}_{\eta\mu}}
-\pfrac{X^\eta}{x^\nu}\pfrac{X^\mu}{x^\beta}\ppfrac{x^\alpha}{X^\eta}{X^\mu}.
\end{equation}
The quantity $\gamma\indices{^{\alpha}_{\nu\beta}}$ defined by Eqs.~(\ref{def:gammainomega2}) is thus the natural choice for an affine connection.
Moreover, this definition implies ``metric compatibility'':
\begin{equation*}
g_{\mu\nu;\alpha}=e\indices{_{\mu}^i}\,\eta_{ij;\alpha}\,e\indices{^j_{\nu}}=0,
\end{equation*}
and also ensures that the mixed representation $R\indices{^n_{m\alpha\beta}}$ of the Riemann-Cartan tensor
is equivalent to its full metric representation $R\indices{^\nu_{\mu\alpha\beta}}$.\footnote{%
At this point it becomes evident that adding non-metricity as a dynamical field is only reasonable with a non-vanishing symmetric (tensor) portion of the spin connection.}
The partial derivatives of the tetrads $e\indices{^i_\mu}$ and $e\indices{^\mu_i}$ can then be expressed
by the spin connection $\ho\indices{^j_{i\nu}}$ and the affine connection $\gamma\indices{^\mu_{\xi\nu}}$ as:
\begin{equation*}
\pfrac{e\indices{^i_\mu}}{x^\nu}=-\ho\indices{^i_{j\nu}}\,e\indices{^j_\mu}
+e\indices{^i_\xi}\gamma\indices{^\xi_{\mu\nu}},\qquad
\pfrac{e\indices{^\mu_i}}{x^\nu}=e\indices{^\mu_j}\,\ho\indices{^j_{i\nu}}-\gamma\indices{^\mu_{\xi\nu}}e\indices{^\xi_i}.
\end{equation*}
Consequently, the canonical field equation~(\ref{eq:e-deri2}) acquires the form
\begin{align}
\pfrac{\tilde{\HCd}_\mathrm{Gr}}{\tilde{k}\indices{_i^{\mu\nu}}}&=\onehalf\left(\pfrac{e\indices{^i_\mu}}{x^\nu}
+\ho\indices{^i_{j\nu}}\,e\indices{^j_\mu}-\pfrac{e\indices{^i_\nu}}{x^\mu}-\ho\indices{^i_{j\mu}}\,e\indices{^j_\nu}\right)\nonumber\\
&=\onehalf e\indices{^i_\xi}\left(\gamma\indices{^\xi_{\mu\nu}}-\gamma\indices{^\xi_{\nu\mu}}\right)
=e\indices{^i_\xi}\,s\indices{^\xi_{\mu\nu}}\equiv s\indices{^i_{\mu\nu}}.
\label{eq:k-deri-metr}
\end{align}
Hereby the skew-symmetric part of the affine connection,
\begin{equation}\label{def:torsion}
s\indices{^\xi_{\mu\nu}}=\onehalf\left(\gamma\indices{^\xi_{\mu\nu}}-\gamma\indices{^\xi_{\nu\mu}}\right)
\end{equation}
is identified with the Cartan torsion tensor.

These relations allow now to re-write the components of the generalized Dirac equation with partial derivatives of the tetrads as:
\begin{align*}
\pfrac{e\indices{^\alpha_j}}{x^\alpha}&-e\indices{^\alpha_n}\ho\indices{^n_{j\alpha}}
-e\indices{^\alpha_j}e\indices{^n_\xi}\pfrac{e\indices{^\xi_n}}{x^\alpha}\\
&=-\gamma\indices{^\alpha_{\xi\alpha}}e\indices{^\xi_j}-e\indices{^\alpha_j}e\indices{^n_\xi}
\left(e\indices{^\xi_m}\ho\indices{^m_{n\alpha}}-\gamma\indices{^\xi_{\beta\alpha}}e\indices{^\beta_n}\right)\\
&=-\gamma\indices{^\alpha_{\xi\alpha}}\,e\indices{^\xi_j}-\cancel{e\indices{^\alpha_j}\,\ho\indices{^m_{m\alpha}}}
+\gamma\indices{^\alpha_{\alpha\xi}}\,e\indices{^\xi_j}\\
&=2s\indices{^\alpha_{\alpha\xi}}\,e\indices{^\xi_j}.
\end{align*}
Notice that  here the spin connection term vanishes due to the skew-symmetry in its first index pair.

By the same token, one gets for the first factor in Eq.~(\ref{eq:gen-dirac-final})
\begin{align*}
&\quad\sigma^{ji}\Bigg(\pfrac{e\indices{^\alpha_j}}{x^\alpha}\,e\indices{_i^\beta}
-e\indices{^\alpha_k}\,\ho\indices{^k_{j\alpha}}\,e\indices{_i^\beta}
+e\indices{^\alpha_j}\,\pfrac{e\indices{_i^\beta}}{x^\alpha}\\
&\qquad\quad-e\indices{^\alpha_j}\,e\indices{^\beta_k}\,\ho\indices{^k_{i\alpha}}
-e\indices{^\alpha_j}\,e\indices{_i^\beta}\,e\indices{^k_\xi}\pfrac{e\indices{^\xi_k}}{x^\alpha}\Bigg)\\
&=\sigma^{ji}\left(-\gamma\indices{^\alpha_{\xi\alpha}}\,e\indices{^\xi_j}\,e\indices{_i^\beta}
-\gamma\indices{^\beta_{\xi\alpha}}\,e\indices{^\xi_i}\,e\indices{^\alpha_j}
+\gamma\indices{^\xi_{\xi\alpha}}\,e\indices{_i^\beta}\,e\indices{^\alpha_j}\right)\\
&=e\indices{^\alpha_j}\,\sigma^{ji}\left(2e\indices{_i^\beta}\,s\indices{^\xi_{\xi\alpha}}
-e\indices{_i^\xi}\,s\indices{^\beta_{\xi\alpha}}\right).
\end{align*}
The generalized Dirac equation thus naturally simplifies to:
\begin{align*}
&\;\frac{\rmi}{3M}e\indices{^\alpha_j}\,\sigma^{ji}\Bigg[\bigg(2e\indices{_i^\beta}\,s\indices{^\xi_{\xi\alpha}}
-e\indices{_i^\xi}\,s\indices{^\beta_{\xi\alpha}}\bigg)
\left(\pfrac{\psi}{x^\beta}-\iquarter\ho\indices{^n_{m\beta}}\,\sigma\indices{_n^m}\psi\right)\\
&\qquad\qquad\quad-\iquarter e\indices{_i^\beta}\,\sigma\indices{_n^m}
\left(\pfrac{\ho\indices{^n_{m\beta}}}{x^\alpha}+\ho\indices{^n_{k\alpha}}\,\ho\indices{^k_{m\beta}}\right)\psi\Bigg]\\
&=\rmi\,\gamma^j\,e\indices{_j^\beta}\left(\pfrac{\psi}{x^\beta}-\iquarter\ho\indices{^n_{m\beta}}\,\sigma\indices{_n^m}\,\psi
+s\indices{^\xi_{\xi\beta}}\psi\right)-m\,\psi,
\end{align*}
or with the abbreviations $\gamma^\beta\equiv\gamma^j\,e\indices{_j^\beta}$ and $\sigma^{\alpha\beta}\equiv e\indices{^\alpha_j}\,\sigma^{ji}e\indices{_i^\beta}$:
\begin{align}
&\,\left[\rmi\gamma^\beta\left(\pfrac{}{x^\beta}-\iquarter\ho\indices{^n_{m\beta}}\,\sigma\indices{_n^m}+s\indices{^\xi_{\xi\beta}}\right)-m\right]\psi\label{eq:gen-dirac-mc}\\
=&\,\frac{1}{3M}\bigg[\frac{1}{8}\,\sigma^{\alpha\beta}\,\sigma\indices{_n^m}\,R\indices{^n_{m\alpha\beta}}\nonumber\\
&\qquad-\rmi\left(2s\indices{^\xi_{\xi\beta}}\,\sigma^{\beta\nu}+s\indices{^\nu_{\eta\beta}}\,\sigma^{\beta\eta}\right)
\left(\pfrac{}{x^\nu}-\iquarter\ho\indices{^n_{m\nu}}\,\sigma\indices{_n^m}\right)\bigg]\psi.\nonumber
\end{align}
This shows that the ``minimal coupling'' prescriptions emerges naturally,
and, moreover, that the Dirac particle couples directly to the Riemann-Cartan curvature tensor,
with coupling constant proportional to $M^{-1}$.
\subsection{Generalized Dirac equation with zero torsion}
Neglecting torsion, equation~(\ref{eq:gen-dirac-mc}) further simplifies to:
\begin{equation}\label{eq:gen-dirac-mc-2}
\rmi\gamma^\beta\left(\pfrac{\psi}{x^\beta}-\iquarter\ho\indices{^n_{m\beta}}\sigma\indices{_n^m}\psi\right)
-\left(m+\frac{\sigma^{\alpha\beta}\sigma\indices{_n^m}}{24M}R\indices{^n_{m\alpha\beta}}\right)\psi=0.
\end{equation}
The contraction of the Riemann-Cartan tensor with the $\sigma$ matrices is shown in App.~\ref{app:riem-contr}
to reduce for the case of zero torsion to twice the Ricci scalar $R$ times the unit matrix in the spinor indices, which finally yields:
\begin{equation}\label{eq:dirac-final-notorsion}\boxed{%
\rmi\,\gamma^\beta\left(\pfrac{\psi}{x^\beta}-\iquarter\ho\indices{^n_{m\beta}}\,\sigma\indices{_n^m}\,\psi\right)-\left(m+\frac{R}{12M}\right)\psi=0.}
\end{equation}
One thus encounters in the Dirac equation a curvature-dependent mass correction term due to a direct interaction of $\psi$ with the gravitational field.
The strength of that interaction is determined by the parameter $M$ that emerges due to the enforcement
of the diffeomorphism invariance and simultaneously by the non-degeneracy of the Hamiltonian $\tilde{\HCd}_{\mathrm{D}}$.
The physical implications of this novel ``spin-gravity coupling mechanism'' will have measurable consequences
in scenarios with $R\gg0$, e.g.\ inflation or neutron star mergers~\citep{Benisty:2019jqz}.
\subsection{The action integral of fermions in dynamical space-time}
Inserting finally Eqs.~(\ref{eq:omega-deri}) for the curvature and~(\ref{def:torsion}) for the torsion
of space-time into the action integral~(\ref{action-integral5}) gives the expression:
\allowdisplaybreaks[0]
\begin{align*}
S_0&=\!\int_{V}\!\Bigg[
\tilde{\kappabar}^\nu\!\left(\pfrac{\psi}{x^\nu}-\iquarter\ho\indices{^i_{j\nu}}\,\sigma\indices{_i^j}\,\psi\right)
+\left(\pfrac{\psibar}{x^\nu}+\iquarter\ho\indices{^i_{j\nu}}\,\psibar\,\sigma\indices{_i^j}\right)\tilde{\kappa}^\nu\nonumber\\
&+\tilde{k}\indices{_i^{\mu\nu}}s\indices{^i_{\mu\nu}}\!
+\frac{1}{2}\hoc\indices{_i^{j\mu\nu}}R\indices{^i_{j\mu\nu}}\!-\tilde{\HCd}_{\mathrm{D}}\!-
\tilde{\HCd}_{\mathrm{Gr}}\big(\tilde{k}\indices{_i^{\mu\nu}},\hoc\indices{_i^{j\mu\nu}},e\indices{^i_\mu}\big)\Bigg]\,\dd^4x.
\end{align*}
This makes obvious that:
\begin{enumerate}
\item By the gauge process, the originally non-covariant partial derivatives of the spinor field are converted into the covariant derivatives
via coupling to the connection $\ho\indices{^i_{j\nu}}$.
For the dynamical geometry, the (non-covariant) derivatives of the connection and tetrad are promoted to the (covariant) Riemann-Cartan curvature
and torsion tensors, respectively.
\item The functional dependence of the free gravity Hamiltonian $\tilde{\HCd}_{\mathrm{Gr}}$ on the momentum fields
$\tilde{k}\indices{_i^{\mu\nu}}$ and $\hoc\indices{_i^{j\mu\nu}}$ must be postulated based on physical reasoning on the structure of space-time.
The coupling to the source field $\psi,\psibar$ then determines the dynamics of the system of the fermion field and space-time.
\end{enumerate}
If in particular the free gravity Hamiltonian $\tilde{\HCd}_{\mathrm{Gr}}$ does not depend on the momentum $\tilde{k}\indices{_i^{\mu\nu}}$,
then the variation of the action leads to $s\indices{^\xi_{\mu\nu}}=0$, i.e., a torsion-free space-time according to Eq.~(\ref{eq:k-deri-metr}).
Similarly, if $\tilde{\HCd}_{\mathrm{Gr}}$ does not depend on the momentum $\hoc\indices{_i^{j\mu\nu}}$,
we then describe a flat geometry as $R\indices{^i_{j\mu\nu}}=0$ follows as the corresponding field equation~(\ref{eq:omega-deri}).
\section{Free gravitational Hamiltonian}\label{sec:free-grav-ham}
Similar to the Hamiltonians $\tilde{\HCd}_{\mathrm{D}}$ of the free fermion system,
the Lagrangian resp.\ Hamiltonian $\tilde{\HCd}_{\mathrm{Gr}}$ of the free (uncoupled) gauge field---the
gravitational field---must be set up on the basis of physical reasoning in conjunction with appropriate physical measurements.
The canonical gauge procedure merely determines the coupling of the fields in question
by requiring the combined system to be diffeomorphism-invariant.
Similar to all other field theories, $\tilde{\HCd}_{\mathrm{Gr}}$ must be quadratic
in the momentum fields $\tilde{q}\indices{_{j}^{k\beta\alpha}}$ and $\tilde{k}\indices{_{j}^{\beta\alpha}}$
in order to encounter well-defined duality relations of momenta and corresponding ``velocities''~\citep{benisty18c}.
With quadratic momentum dependence, the Riemann-Cartan curvature and the torsion become ``propagating''
field strengths, associated with the respective connection and tetrad fields.
A reasonable choice is thus to postulate $\tilde{\HCd}_{\mathrm{Gr,post}}(\tilde{q},\tilde{k},e)$ as
\begin{align}
\tilde{\HCd}_{\mathrm{Gr,post}}&=\frac{1}{4\,g_{1}\dete}\,\tilde{q}\indices{_{i}^{j\alpha\beta}}
\tilde{q}\indices{_{j}^{i\xi\lambda}}\,g_{\alpha\xi}g_{\beta\lambda}+
g_2\,\tilde{q}\indices{_{i}^{j\alpha\beta}}e\indices{^i_\alpha}e\indices{^n_\beta}\,\eta_{nj}\nonumber\\
&\quad+\frac{g_{3}}{2\dete}\,\tilde{k}\indices{_{i}^{\alpha\beta}}
\tilde{k}\indices{_{j}^{\xi\lambda}}\,\eta^{ij}\,g_{\alpha\xi}\,g_{\beta\lambda},
\quad g_{\alpha\beta}=\eta_{ij}e\indices{^i_\alpha}e\indices{^j_\beta}.
\label{eq:ham-free-grav}
\end{align}
$g_1$, $g_2$, and $g_3$ are coupling constants, which must be adapted to measurements/experiments.
For the particular choice $g_3=0$, the resulting field equation is satisfied by the Schwarzschild
and the more general Kerr metric~\citep{struckmeier17a,stephenson58,kehm17}.
\section{Summary and outlook}\label{sec:conclusions}
Based on the evident postulate that the description of physics should be the same in any coordinate frame,
we have derived the closed set of canonical field equations describing the mutual interaction of spinors with a gravitational field.
The first precondition for this unified description is the knowledge of the free (uncoupled) dynamics of the spinors,
which was assumed to be given by a non-degenerate Dirac Lagrangian or its equivalent Hamiltonian counterpart.
As S.~Gasiorovicz~\citep{gasiorowicz66} noted, the quadratic velocity term in the Dirac
Lagrangian~(\ref{eq:ld-dirac})---which renders it non-degenerate without changing the resulting Dirac
equation---does ``not appear to be necessary'', but ``cannot be logically excluded''.
As a consequence, we recover the minimal coupling scheme of spinors to curved space-time, and in addition a new Fermi-like
interaction term in the Dirac equation~(\ref{eq:dirac-final-notorsion}) leading to an anomalous mass correction.
Its strength is determined by a new mass (or length) parameter that inevitably arises in a non-degenerate Hamiltonian for dimensional reasons.
While unnecessary for free spinors, this parameter becomes a physical quantity in curved space-time geometries
that will fundamentally modify cosmological \citep{vasak20,Benisty:2019jqz} and astrophysical models.

The second precondition for a unified description is the knowledge of the Lagrangian
resp.\ Hamiltonian of the free (uncoupled) dynamics of the gravitational field.
Here, its functional dependence on the momentum fields determines via the canonical equations the dynamical space-time structure.
A non-degenerate gravity Hamiltonian~(\ref{eq:ham-free-grav}) was previously discussed in~\citep{struckmeier17a} for the case of vanishing torsion.
As the corresponding field equations are satisfied not only by the Schwarzschild metric~\citep{stephenson58},
but also by the more general Schwarzschild-De~Sitter and the Kerr-De~Sitter metrics, this ``free gravity'' Hamiltonian is consistent with actual measurements.
Its cosmological implications have been scrutinized in Refs.~\citep{vasak20} and~\citep{vasak20b}.

On the basis of these ``free'' Hamiltonians, we have derived the unified dynamics of spinors and space-time
by a coupled set of eight first-order field equations following the line of non-Abelian gauge theories via the canonical transformation formalism.
The equations include torsion of space-time and describe a new direct coupling of spinors with the Riemann curvature tensor.
Their solutions are therefore expected to differ from those of the Einstein equation.
The consequences for cosmological models must be clarified in future studies.
\subsection*{Acknowledgments}
The authors are indebted to the ``Walter Greiner-Gesellschaft e.V.'' in Frankfurt for its support.
DV is grateful for support from the Fueck Foundation.
We thank F.W.~Hehl (Universit\"at K\"oln) for inspiring discussions.

\appendix
\section{Explicit calculations}
\subsection{Calculation of the tetrad field \texorpdfstring{$E\indices{^I_\alpha}$}{E} contribution to Eq.~(\ref{eq:f3derivative2})\label{app:tetrad-calc}}
First of all, we expand the third term in Eq.~(\ref{eq:f3derivative2}),
\begin{equation}\label{eq:f3derivative2exp}
\tilde{k}\indices{_i^{\mu\nu}}\!\pfrac{}{x^\nu}\!\left(\!\Lambda\indices{^i_J}
\pfrac{X^\alpha}{x^\mu}\right)E\indices{^{J}_{\alpha}}\!
=\tilde{k}\indices{_i^{\mu\nu}}\!\left(\pfrac{\Lambda\indices{^i_J}}{x^\nu}\pfrac{X^\alpha}{x^\mu}+
\Lambda\indices{^i_J}\ppfrac{X^\alpha}{x^\mu}{x^\nu}\right)E\indices{^{J}_{\alpha}},
\end{equation}
and re-write the transformation rule~(\ref{eq:conn-trans}) for the gauge field $\ho\indices{^i_{j\nu}}$ as\vspace*{-1mm}
\begin{equation*}
\pfrac{\Lambda\indices{^i_J}}{x^\nu}=\Lambda\indices{^i_I}\,\HO\indices{^I_{J\xi}}\,\pfrac{X^\xi}{x^\nu}-
\ho\indices{^i_{j\nu}}\,\Lambda\indices{^j_J}.
\end{equation*}
With the canonical transformation rule~(\ref{eq:tet-trans}) written in the equivalent form\vspace*{-1mm}
\begin{equation*}
\pfrac{X^\alpha}{x^\mu}=E\indices{^\alpha_I}\,\Lambda\indices{^I_i}\,e\indices{^i_\mu},
\end{equation*}
we find for the $x^\nu$-derivative
\begin{align}
\ppfrac{X^\alpha}{x^\mu}{x^\nu}
&=\pfrac{E\indices{^\alpha_I}}{x^\nu}\Lambda\indices{^I_i}e\indices{^i_\mu}
+E\indices{^\alpha_I}\Lambda\indices{^I_i}\pfrac{e\indices{^i_\mu}}{x^\nu}
+E\indices{^\alpha_I}\pfrac{\Lambda\indices{^I_j}}{x^\nu}\,e\indices{^j_\mu}\nonumber\\
&=\pfrac{E\indices{^\alpha_I}}{x^\nu}E\indices{^I_\eta}\pfrac{X^\eta}{x^\mu}
+\pfrac{X^\alpha}{x^\xi}e\indices{^\xi_i}\pfrac{e\indices{^i_\mu}}{x^\nu}\nonumber\\
&\quad+E\indices{^\alpha_I}\left(\Lambda\indices{^I_i}\ho\indices{^i_{j\nu}}
-\HO\indices{^I_{J\xi}}\Lambda\indices{^J_j}\pfrac{X^\xi}{x^\nu}\right)e\indices{^j_\mu}\nonumber\\
&=-E\indices{^\alpha_I}\pfrac{E\indices{^I_\eta}}{X^\xi}\pfrac{X^\xi}{x^\nu}\pfrac{X^\eta}{x^\mu}
+e\indices{^\xi_i}\,\pfrac{e\indices{^i_\mu}}{x^\nu}\,\pfrac{X^\alpha}{x^\xi}\nonumber\\
&\quad+\pfrac{X^\alpha}{x^\xi}\,e\indices{^\xi_i}\,\ho\indices{^i_{j\nu}}\,e\indices{^j_\mu}-
E\indices{^\alpha_I}\,\HO\indices{^I_{J\xi}}\,E\indices{^J_\eta}\,\pfrac{X^\eta}{x^\mu}\pfrac{X^\xi}{x^\nu}\nonumber\\
&=e\indices{^\xi_i}\left(\pfrac{e\indices{^i_\mu}}{x^\nu}+\ho\indices{^i_{j\nu}}e\indices{^j_\mu}\right)\pfrac{X^\alpha}{x^\xi}\nonumber\\
&\quad-E\indices{^\alpha_I}\left(\pfrac{E\indices{^I_\eta}}{X^\xi}+\HO\indices{^I_{J\xi}}E\indices{^J_\eta}\right)
\pfrac{X^\eta}{x^\mu}\pfrac{X^\xi}{x^\nu}.
\label{eq:X2-trans}
\end{align}
Inserting now Eq.~(\ref{eq:X2-trans}) into Eq.~(\ref{eq:f3derivative2exp}), one finds
\begin{align*}
&\quad-\tilde{k}\indices{_i^{\mu\nu}}\,\pfrac{}{x^\nu}\left(\Lambda\indices{^i_J}
\pfrac{X^\alpha}{x^\mu}\right)E\indices{^{J}_{\alpha}}\\
&=-\tilde{k}\indices{_i^{\mu\nu}}\pfrac{\Lambda\indices{^i_J}}{x^\nu}\pfrac{X^\alpha}{x^\mu}E\indices{^{J}_{\alpha}}-
\tilde{k}\indices{_i^{(\mu\nu)}}\Lambda\indices{^i_J}\ppfrac{X^\alpha}{x^\mu}{x^\nu}E\indices{^{J}_{\alpha}}\\
&=-\tilde{k}\indices{_i^{\mu\nu}}\Bigg(\Lambda\indices{^i_I}\,\HO\indices{^I_{J\xi}}\pfrac{X^\xi}{x^\nu}-
\ho\indices{^i_{j\nu}}\Lambda\indices{^j_J}\Bigg)\,E\indices{^{J}_{\alpha}}\pfrac{X^\alpha}{x^\mu}\\
&\quad-\tilde{k}\indices{_i^{(\mu\nu)}}\Lambda\indices{^i_J}E\indices{^{J}_{\alpha}}\ppfrac{X^\alpha}{x^\mu}{x^\nu}\\
&=-\tilde{k}\indices{_i^{\mu\nu}}\Lambda\indices{^i_I}\HO\indices{^I_{J\xi}}E\indices{^{J}_{\alpha}}
\pfrac{X^\alpha}{x^\mu}\pfrac{X^\xi}{x^\nu}+\tilde{k}\indices{_i^{\mu\nu}}
\ho\indices{^i_{j\nu}}\Lambda\indices{^j_J}E\indices{^{J}_{\alpha}}\pfrac{X^\alpha}{x^\mu}\\
&\quad+\onehalf\tilde{k}\indices{_i^{\mu\nu}}\Lambda\indices{^i_I}\pfrac{X^\xi}{x^\nu}\pfrac{X^\eta}{x^\mu}
\Bigg(\pfrac{E\indices{^I_\eta}}{X^\xi}+\pfrac{E\indices{^I_\xi}}{X^\eta}\!+\HO\indices{^I_{J\xi}}
E\indices{^J_\eta}\!+\HO\indices{^I_{J\eta}}E\indices{^J_\xi}\!\Bigg)\nonumber\\
&\quad-\onehalf\tilde{k}\indices{_n^{\mu\nu}}
\Lambda\indices{^n_J}E\indices{^{J}_{\alpha}}\pfrac{X^\alpha}{x^\xi}e\indices{^\xi_i}\!\left(\pfrac{e\indices{^i_\mu}}{x^\nu}+
\pfrac{e\indices{^i_\nu}}{x^\mu}+\ho\indices{^i_{j\nu}}e\indices{^j_\mu}\!+\ho\indices{^i_{j\mu}}e\indices{^j_\nu}\!\right)\\
&=\tilde{k}\indices{_i^{\mu\nu}}\ho\indices{^i_{j\nu}}e\indices{^{j}_{\mu}}
-\tilde{K}\indices{_I^{\eta\xi}}\HO\indices{^I_{J\xi}}E\indices{^{J}_{\eta}}\detpartial{X}{x}\\
&\quad-\onehalf\tilde{k}\indices{_i^{\mu\nu}}\left(\pfrac{e\indices{^i_\mu}}{x^\nu}+\pfrac{e\indices{^i_\nu}}{x^\mu}
+\ho\indices{^i_{j\nu}}e\indices{^j_\mu}+\ho\indices{^i_{j\mu}}e\indices{^j_\nu}\right)\\
&\quad+\onehalf\tilde{K}\indices{_I^{\eta\xi}}\Bigg(\pfrac{E\indices{^I_\eta}}{X^\xi}+\pfrac{E\indices{^I_\xi}}{X^\eta}
+\HO\indices{^I_{J\xi}}E\indices{^J_\eta}+\HO\indices{^I_{J\eta}}E\indices{^J_\xi}\Bigg)\detpartial{X}{x},
\end{align*}
hence finally
\begin{align*}
&\quad\,\tilde{k}\indices{_i^{\mu\nu}}\,\pfrac{}{x^\nu}\left(\Lambda\indices{^i_J}
\pfrac{X^\alpha}{x^\mu}\right)E\indices{^{J}_{\alpha}}\nonumber\\
&=\onehalf\tilde{k}\indices{_i^{\mu\nu}}\left(\pfrac{e\indices{^{i}_{\mu}}}{x^\nu}+\pfrac{e\indices{^{i}_{\nu}}}{x^\mu}-
\ho\indices{^i_{j\nu}}e\indices{^{j}_{\mu}}+\ho\indices{^i_{j\mu}}e\indices{^{j}_{\nu}}\right)\nonumber\\
&\quad-\onehalf\tilde{K}\indices{_I^{\mu\nu}}\Bigg(\pfrac{E\indices{^{I}_{\mu}}}{X^\nu}+\pfrac{E\indices{^{I}_{\nu}}}{X^\mu}-
\HO\indices{^I_{J\nu}}E\indices{^{J}_{\mu}}+\HO\indices{^I_{J\mu}}E\indices{^{J}_{\nu}}\Bigg)\detpartial{X}{x}.
\end{align*}
\subsection{Proving the equivalence of the transformation rules of
the affine and spin connections\label{app:gamma_connection}}
In order to proof that $\gamma\indices{^\eta_{\mu\nu}}$---as defined in Eq.~(\ref{def:gammainomega2})---%
transforms according to Eq.~(\ref{gammatransforminverse}) and hence represents the affine connection, we recall the
transformation laws for the tetrad (Eq.~(\ref{eq:tet-trans})) and for the spin connection from Eq.~(\ref{eq:omegatransform1}):
\begin{equation*}
e\indices{^i_\mu}=\Lambda\indices{^i_I}E\indices{^I_\alpha}\pfrac{X^\alpha}{x^\mu},\quad
\ho\indices{^i_{j\nu}}=\Lambda\indices{^i_I}\,\HO\indices{^I_{J\alpha}}\Lambda\indices{^J_j}\pfrac{X^\alpha}{x^\nu}
+\Lambda\indices{^i_I}\pfrac{\Lambda\indices{^I_j}}{x^\nu}.
\end{equation*}
Then
\begin{align*}
&\gamma\indices{^\eta_{\mu\nu}}
=e\indices{^\eta_i}\ho\indices{^i_{j\nu}}e\indices{^j_\mu}+e\indices{^\eta_i}\pfrac{e\indices{^i_\mu}}{x^{\nu}}\\
&=e\indices{^\eta_i}\Bigg[\Lambda\indices{^i_I}\Bigg(\HO\indices{^I_{J\alpha}}\Lambda\indices{^J_j}\pfrac{X^\alpha}{x^\nu}
+\pfrac{\Lambda\indices{^I_j}}{x^\nu}\Bigg)\,e\indices{^j_\mu}\!
+\pfrac{}{x^{\nu}}\left(\Lambda\indices{^i_I}E\indices{^I_\alpha}\pfrac{X^\alpha}{x^\mu}\right)\!\Bigg].
\end{align*}\onecolumn\noindent
By virtue of the identity
\begin{equation*}
e\indices{^\eta_i}\pfrac{\Lambda\indices{^i_I}}{x^{\nu}}E\indices{^I_\alpha}\pfrac{X^\alpha}{x^\mu}
=e\indices{^\eta_i}\pfrac{\Lambda\indices{^i_I}}{x^{\nu}}\Lambda\indices{^I_j}e\indices{^j_\mu}
=-e\indices{^\eta_i}\Lambda\indices{^i_I}\pfrac{\Lambda\indices{^I_j}}{x^{\nu}}e\indices{^j_\mu},
\end{equation*}
the corresponding two terms cancel, hence
\begin{equation*}
\gamma\indices{^\eta_{\mu\nu}}
\!=e\indices{^\eta_i}\Lambda\indices{^i_I}\!\left(\pfrac{E\indices{^I_\alpha}}{x^{\nu}}\pfrac{X^\alpha}{x^\mu}
+E\indices{^I_\alpha}\ppfrac{X^\alpha}{x^\mu}{x^\nu}
+\HO\indices{^I_{J\alpha}}\Lambda\indices{^J_j}e\indices{^j_\mu}\pfrac{X^\alpha}{x^\nu}\right).
\end{equation*}
With
\begin{equation*}
e\indices{^\eta_i}\Lambda\indices{^i_I}=E\indices{^\xi_I}\pfrac{x^\eta}{X^\xi},
\end{equation*}
this yields
\begin{align*}
\gamma\indices{^\eta_{\mu\nu}}\!&=E\indices{^\xi_I}\left(\pfrac{E\indices{^I_\alpha}}{x^{\nu}}\pfrac{X^\alpha}{x^\mu}
+E\indices{^I_\alpha}\ppfrac{X^\alpha}{x^\mu}{x^\nu}
+\;\HO\indices{^I_{J\alpha}}E\indices{^J_\beta}\pfrac{X^\beta}{x^\mu}\pfrac{X^\alpha}{x^\nu}\right)\pfrac{x^\eta}{X^\xi}\\
&=E\indices{^\xi_I}\!\left(\pfrac{E\indices{^I_\alpha}}{X^{\beta}}
+\HO\indices{^I_{J\beta}}E\indices{^J_\alpha}\!\right)\pfrac{X^\beta}{x^\nu}\pfrac{X^\alpha}{x^\mu}\pfrac{x^\eta}{X^\xi}
\!+\!\ppfrac{X^\xi}{x^\mu}{x^\nu}\pfrac{x^\eta}{X^\xi}.
\end{align*}
With the definition of $\Gamma\indices{^\eta_{\mu\nu}}$ corresponding to that of $\gamma\indices{^\eta_{\mu\nu}}$,
\begin{equation*}
\Gamma\indices{^\eta_{\mu\nu}}
=E\indices{^\eta_I}\HO\indices{^I_{J\nu}}E\indices{^J_\mu}+E\indices{^\eta_I}\pfrac{E\indices{^I_\mu}}{X^{\nu}},
\end{equation*}
the transformation law for $\gamma\indices{^\eta_{\mu\nu}}$ finally emerges as
\begin{equation*}
\gamma\indices{^\eta_{\mu\nu}}=\Gamma\indices{^\xi_{\alpha\beta}}\pfrac{X^\beta}{x^\nu}\pfrac{X^\alpha}{x^\mu}\pfrac{x^\eta}{X^\xi}
+\ppfrac{X^\xi}{x^\mu}{x^\nu}\pfrac{x^\eta}{X^\xi}.
\end{equation*}
The other direction of the proof, namely the derivation of the transformation
relation~(\ref{gammatransforminverse}) from Eq.~(\ref{eq:omegatransform1}), is obvious.

In a similar way it is straightforward to prove that the tensor
$R\indices{^i_{j\nu\mu}}$, defined in Eq.~(\ref{eq:omega-deri}), is the mixed representation
of the Riemann-Cartan tensor $R\indices{^\alpha_{\beta\nu\mu}}$ by inserting
\begin{equation*}
\ho\indices{^i_{j\nu}}=e\indices{^i_\beta}\,\gamma\indices{^\beta_{\alpha\nu}}\,e\indices{^\alpha_j}-\pfrac{e\indices{^i_\alpha}}{x^\nu}\,e\indices{^\alpha_j}.
\end{equation*}
\subsection{Calculation of the gauge field \texorpdfstring{$\HO\indices{^I_{J\alpha}}$}{[Omega]} contribution to Eq.~(\ref{eq:f3derivative2})\label{app:gauge-calc}}
In order to express all coefficients in the term proportional to $\hoc\indices{_i^{j\mu\nu}}$ of Eq.~(\ref{eq:f3derivative2}),
we first write this term in expanded form:
\begin{align*}
&\quad-\hoc\indices{_i^{j\mu\nu}}\Bigg[\HO\indices{^I_{J\alpha}}\,\pfrac{}{x^\nu}
\left(\Lambda\indices{^i_I}\,\Lambda\indices{^J_j}\,\pfrac{X^\alpha}{x^\mu} \right)
+\pfrac{}{x^\nu}\Bigg(\Lambda\indices{^i_I}\,\pfrac{\Lambda\indices{^I_j}}{x^\mu}\Bigg)\Bigg]\\
&=-\hoc\indices{_i^{j\mu\nu}} \HO\indices{^I_{J\alpha}}
\Bigg( \pfrac{\Lambda\indices{^i_I}}{x^\nu}\,\Lambda\indices{^J_j}\,\pfrac{X^\alpha}{x^\mu}+
\Lambda\indices{^i_I}\,\pfrac{\Lambda\indices{^J_j}}{x^\nu}\,\pfrac{X^\alpha}{x^\mu}+
\Lambda\indices{^i_I}\,\Lambda\indices{^J_j}\,\ppfrac{X^\alpha}{x^\mu}{x^\nu}\Bigg)
-\hoc\indices{_i^{j\mu\nu}}\Bigg(\pfrac{\Lambda\indices{^i_I}}{x^\nu}\,\pfrac{\Lambda\indices{^I_j}}{x^\mu}+
\Lambda\indices{^i_I}\ppfrac{\Lambda\indices{^I_j}}{x^\mu}{x^\nu}\Bigg).
\end{align*}
With the transformation rule~(\ref{eq:omegatransform1}) solved for $\HO\indices{^I_{J\alpha}}$
\begin{equation*}
\HO\indices{^I_{J\alpha}}=\left(\Lambda\indices{^I_n}\,\ho\indices{^n_{m\xi}}
-\pfrac{\Lambda\indices{^I_m}}{x^\xi}\right)\Lambda\indices{^m_J}\pfrac{x^\xi}{X^\alpha},
\end{equation*}
Eq.~(\ref{eq:f3derivative3}) is expressed equivalently in terms of the original fields as
\begin{align}
&-\hoc\indices{_i^{j\mu\nu}}\Bigg[\left(\Lambda\indices{^I_n}\ho\indices{^n_{m\xi}}
-\pfrac{\Lambda\indices{^I_m}}{x^\xi}\right)\Bigg(\pfrac{\Lambda\indices{^i_I}}{x^\nu}\delta^m_j\delta^\xi_\mu+
\Lambda\indices{^i_I}\pfrac{\Lambda\indices{^J_j}}{x^\nu}\Lambda\indices{^m_J}\delta^\xi_\mu+
\Lambda\indices{^i_I}\delta^m_j\ppfrac{X^\alpha}{x^\mu}{x^\nu}\pfrac{x^\xi}{X^\alpha}\Bigg)
+\pfrac{\Lambda\indices{^i_I}}{x^\nu}\,\pfrac{\Lambda\indices{^I_j}}{x^\mu}+
\Lambda\indices{^i_I}\ppfrac{\Lambda\indices{^I_j}}{x^\mu}{x^\nu}\Bigg]\nonumber\\
&=-\hoc\indices{_i^{j\mu\nu}}\Bigg(\ho\indices{^n_{j\mu}}\Lambda\indices{^I_n}\pfrac{\Lambda\indices{^i_I}}{x^\nu}
+\ho\indices{^i_{n\mu}}\Lambda\indices{^n_J}\pfrac{\Lambda\indices{^J_j}}{x^\nu}
+\ho\indices{^i_{j\xi}}\ppfrac{X^\alpha}{x^\mu}{x^\nu}\pfrac{x^\xi}{X^\alpha}
-\cancel{\pfrac{\Lambda\indices{^I_j}}{x^\mu}\pfrac{\Lambda\indices{^i_I}}{x^\nu}}\nonumber\\
&\quad-\,\Lambda\indices{^i_I}\pfrac{\Lambda\indices{^I_n}}{x^\mu}\Lambda\indices{^n_J}\pfrac{\Lambda\indices{^J_j}}{x^\nu}
-\Lambda\indices{^i_I}\pfrac{\Lambda\indices{^I_j}}{x^\xi}\ppfrac{X^\alpha}{x^\mu}{x^\nu}\pfrac{x^\xi}{X^\alpha}
+\cancel{\pfrac{\Lambda\indices{^i_I}}{x^\nu}\pfrac{\Lambda\indices{^I_j}}{x^\mu}}
+\Lambda\indices{^i_I}\ppfrac{\Lambda\indices{^I_j}}{x^\mu}{x^\nu}\Bigg)\nonumber\\
&=-\hoc\indices{_i^{j\mu\nu}}\Bigg[\ho\indices{^i_{n\mu}}\Lambda\indices{^n_J}\pfrac{\Lambda\indices{^J_j}}{x^\nu}
-\ho\indices{^n_{j\mu}}\Lambda\indices{^i_I}\pfrac{\Lambda\indices{^I_n}}{x^\nu}
+\Bigg(\ho\indices{^i_{j\xi}}-\Lambda\indices{^i_I}\pfrac{\Lambda\indices{^I_j}}{x^\xi}\Bigg)
\ppfrac{X^\alpha}{x^\mu}{x^\nu}\pfrac{x^\xi}{X^\alpha}
-\Lambda\indices{^i_I}\pfrac{\Lambda\indices{^I_n}}{x^\mu}
\Lambda\indices{^n_J}\pfrac{\Lambda\indices{^J_j}}{x^\nu}
+\Lambda\indices{^i_I}\ppfrac{\Lambda\indices{^I_j}}{x^\mu}{x^\nu}\Bigg].\label{F21derivative4}
\end{align}
Now, the transformation rule~(\ref{eq:omegatransform1}) is inserted in the form
\begin{equation*}
\Lambda\indices{^i_I}\,\pfrac{\Lambda\indices{^I_j}}{x^\nu}=
\ho\indices{^i_{j\nu}}-\Lambda\indices{^i_I}\,\HO\indices{^I_{J\alpha}}\Lambda\indices{^J_j}\pfrac{X^\alpha}{x^\nu}
\end{equation*}
and its derivative
\begin{align*}
\Lambda\indices{^i_I}\,\ppfrac{\Lambda\indices{^I_j}}{x^\mu}{x^\nu}&=
\onehalf\Lambda\indices{^i_I}\left(\pfrac{\Lambda\indices{^I_n}}{x^\nu}\ho\indices{^n_{j\mu}}+
\pfrac{\Lambda\indices{^I_n}}{x^\mu}\ho\indices{^n_{j\nu}}\right)+
\onehalf\Bigg(\pfrac{\ho\indices{^i_{j\mu}}}{x^\nu}+\pfrac{\ho\indices{^i_{j\nu}}}{x^\mu}\Bigg)
-\onehalf\Lambda\indices{^i_I}\Bigg(\pfrac{\HO\indices{^I_{J\xi}}}{X^\alpha}+\pfrac{\HO\indices{^I_{J\alpha}}}{X^\xi}\Bigg)\,
\Lambda\indices{^J_j}\pfrac{X^\xi}{x^\mu}\pfrac{X^\alpha}{x^\nu}\\
&\quad-\onehalf\Lambda\indices{^i_I}\HO\indices{^I_{J\alpha}}\Bigg(
\pfrac{\Lambda\indices{^J_j}}{x^\nu}\pfrac{X^\alpha}{x^\mu}+\pfrac{\Lambda\indices{^J_j}}{x^\mu}\pfrac{X^\alpha}{x^\nu}\Bigg)
-\Lambda\indices{^i_I}\,\HO\indices{^I_{J\alpha}}\Lambda\indices{^J_j}\ppfrac{X^\alpha}{x^\mu}{x^\nu}.
\end{align*}
Consolidating all terms yields
\allowdisplaybreaks[0]
\begin{align*}
&-\hoc\indices{_i^{j\mu\nu}}\Bigg[
\onehalf\Bigg(\pfrac{\ho\indices{^i_{j\mu}}}{x^\nu}+\pfrac{\ho\indices{^i_{j\nu}}}{x^\mu}\Bigg)
-\onehalf\Bigg(\pfrac{\HO\indices{^I_{K\beta}}}{X^\alpha}+\pfrac{\HO\indices{^I_{K\alpha}}}{X^\beta}\Bigg)
\Lambda\indices{^i_I}\Lambda\indices{^K_j}\pfrac{X^\beta}{x^\mu}\pfrac{X^\alpha}{x^\nu}\\
&\qquad\quad+\ho\indices{^i_{n\mu}}
\left(\ho\indices{^n_{j\nu}}-\Lambda\indices{^n_I}\,\HO\indices{^I_{J\alpha}}\Lambda\indices{^J_j}\pfrac{X^\alpha}{x^\nu}\right)
-\onehalf\ho\indices{^n_{j\mu}}\left(\ho\indices{^i_{n\nu}}-\Lambda\indices{^i_I}\,\HO\indices{^I_{J\alpha}}
\Lambda\indices{^J_n}\pfrac{X^\alpha}{x^\nu}\right)\nonumber\\
&\qquad\quad+\onehalf\ho\indices{^n_{j\nu}}
\left(\ho\indices{^i_{n\mu}}-\Lambda\indices{^i_I}\,\HO\indices{^I_{J\alpha}}\Lambda\indices{^J_n}\pfrac{X^\alpha}{x^\mu}\right)
+\cancel{\Lambda\indices{^i_I}\,\HO\indices{^I_{J\alpha}}\Lambda\indices{^J_j}\ppfrac{X^\alpha}{x^\mu}{x^\nu}}\\
&\qquad\quad-\,
\left(\ho\indices{^i_{n\mu}}-\Lambda\indices{^i_I}\,\HO\indices{^I_{J\alpha}}\Lambda\indices{^J_n}\pfrac{X^\alpha}{x^\mu}\right)
\left(\ho\indices{^n_{j\nu}}-\Lambda\indices{^n_K}\,\HO\indices{^K_{L\beta}}\Lambda\indices{^L_j}\pfrac{X^\beta}{x^\nu}\right)\\
&\qquad\quad-\onehalf\Lambda\indices{^i_I}\HO\indices{^I_{J\alpha}}\left(
\Lambda\indices{^J_n}\,\ho\indices{^n_{j\nu}}
-\HO\indices{^J_{K\beta}}\Lambda\indices{^K_j}\pfrac{X^\beta}{x^\nu}\right)\pfrac{X^\alpha}{x^\mu}\\
&\qquad\quad-\onehalf\Lambda\indices{^i_I}\HO\indices{^I_{J\alpha}}\left(
\Lambda\indices{^J_n}\,\ho\indices{^n_{j\mu}}
-\HO\indices{^J_{K\beta}}\Lambda\indices{^K_j}\pfrac{X^\beta}{x^\mu}\right)\pfrac{X^\alpha}{x^\nu}
-\cancel{\Lambda\indices{^i_I}\,\HO\indices{^I_{J\alpha}}\Lambda\indices{^J_j}\ppfrac{X^\alpha}{x^\mu}{x^\nu}}\,\Bigg],
\end{align*}
which simplifies, after expanding
\begin{align*}
&-\hoc\indices{_i^{j\mu\nu}}\Bigg[
\onehalf\Bigg(\pfrac{\ho\indices{^i_{j\mu}}}{x^\nu}+\pfrac{\ho\indices{^i_{j\nu}}}{x^\mu}\Bigg)
-\onehalf\Bigg(\pfrac{\HO\indices{^I_{K\beta}}}{X^\alpha}+\pfrac{\HO\indices{^I_{K\alpha}}}{X^\beta}\Bigg)\,
\Lambda\indices{^i_I}\Lambda\indices{^K_j}\pfrac{X^\alpha}{x^\nu}\pfrac{X^\beta}{x^\mu}\\
&\qquad\quad+\ho\indices{^i_{n\mu}}
\ho\indices{^n_{j\nu}}-\ho\indices{^i_{n\mu}}\Lambda\indices{^n_I}\,\HO\indices{^I_{J\alpha}}\Lambda\indices{^J_j}\pfrac{X^\alpha}{x^\nu}
-\onehalf\ho\indices{^n_{j\mu}}\ho\indices{^i_{n\nu}}+\onehalf\ho\indices{^n_{j\mu}}\Lambda\indices{^i_I}\,\HO\indices{^I_{J\alpha}}
\Lambda\indices{^J_n}\pfrac{X^\alpha}{x^\nu}\nonumber\\
&\qquad\quad+\onehalf\ho\indices{^n_{j\nu}}\ho\indices{^i_{n\mu}}-\onehalf\ho\indices{^n_{j\nu}}\Lambda\indices{^i_I}\,
\HO\indices{^I_{J\alpha}}\Lambda\indices{^J_n}\pfrac{X^\alpha}{x^\mu}
-\ho\indices{^i_{n\mu}}\,\ho\indices{^n_{j\nu}}
+\ho\indices{^i_{n\mu}}\Lambda\indices{^n_I}\,\HO\indices{^I_{J\alpha}}\Lambda\indices{^J_j}\pfrac{X^\alpha}{x^\nu}\\
&\qquad\quad+\ho\indices{^n_{j\nu}}\Lambda\indices{^i_I}\,\HO\indices{^I_{J\alpha}}\Lambda\indices{^J_n}\pfrac{X^\alpha}{x^\mu}
-\Lambda\indices{^i_I}\,\HO\indices{^I_{J\alpha}}
\,\HO\indices{^J_{K\beta}}\Lambda\indices{^K_j}\pfrac{X^\alpha}{x^\mu}\pfrac{X^\beta}{x^\nu}\\
&\qquad\quad-\onehalf\Lambda\indices{^i_I}\HO\indices{^I_{J\alpha}}
\Lambda\indices{^J_n}\,\ho\indices{^n_{j\nu}}\pfrac{X^\alpha}{x^\mu}+\onehalf\Lambda\indices{^i_I}\HO\indices{^I_{J\alpha}}
\HO\indices{^J_{K\beta}}\Lambda\indices{^K_j}\pfrac{X^\alpha}{x^\mu}\pfrac{X^\beta}{x^\nu}\\
&\qquad\quad-\onehalf\Lambda\indices{^i_I}\HO\indices{^I_{J\alpha}}
\Lambda\indices{^J_n}\,\ho\indices{^n_{j\mu}}\pfrac{X^\alpha}{x^\nu}+\onehalf\Lambda\indices{^i_I}\HO\indices{^I_{J\alpha}}
\HO\indices{^J_{K\beta}}\Lambda\indices{^K_j}\pfrac{X^\alpha}{x^\nu}\pfrac{X^\beta}{x^\mu}\Bigg]\\
&=-\onehalf\hoc\indices{_i^{j\mu\nu}}\Bigg[
\pfrac{\ho\indices{^i_{j\mu}}}{x^\nu}+\pfrac{\ho\indices{^i_{j\nu}}}{x^\mu}
+\ho\indices{^i_{n\mu}}\,\ho\indices{^n_{j\nu}}-\ho\indices{^i_{n\nu}}\,\ho\indices{^n_{j\mu}}\\
&\qquad\qquad\quad-\Bigg(\pfrac{\HO\indices{^I_{K\alpha}}}{X^\beta}+\pfrac{\HO\indices{^I_{K\beta}}}{X^\alpha}
+\HO\indices{^I_{J\alpha}}\HO\indices{^J_{K\beta}}-\HO\indices{^I_{J\beta}}\HO\indices{^J_{K\alpha}}\Bigg)\,
\Lambda\indices{^i_I}\Lambda\indices{^K_j}\pfrac{X^\alpha}{x^\mu}\pfrac{X^\beta}{x^\nu}\Bigg]\\
&=-\onehalf\hoc\indices{_i^{j\mu\nu}}\Bigg(
\pfrac{\ho\indices{^i_{j\mu}}}{x^\nu}+\pfrac{\ho\indices{^i_{j\nu}}}{x^\mu}
+\ho\indices{^i_{n\mu}}\,\ho\indices{^n_{j\nu}}-\ho\indices{^i_{n\nu}}\,\ho\indices{^n_{j\mu}}\Bigg)\\
&\quad\,+\onehalf\HOc\indices{_I^{J\mu\nu}}\left(\pfrac{\HO\indices{^I_{J\mu}}}{X^\nu}+\pfrac{\HO\indices{^I_{J\nu}}}{X^\mu}
+\HO\indices{^I_{K\mu}}\,\HO\indices{^K_{J\nu}}-\HO\indices{^I_{K\nu}}\,\HO\indices{^K_{J\mu}}\right)\detpartial{X}{x}.
\end{align*}
\subsection{Explicit derivation of the generalized Dirac equation\label{app:gen-dirac}}
\let\bsigma\sigma\let\bgamma\gamma\let\bEins\Eins\relax
In order to derive the generalized Dirac equation for the spinor $\psi$, we eliminate the canonical momentum dependence.
To this end, the first canonical equation~(\ref{eq:can-dirac1b}) is solved for the momentum field $\tilde{\kappa}^\alpha$.
The resulting canonical equation~(\ref{eq:can-dirac1c}) is then inserted into Eq.~(\ref{eq:can-dirac2b}) to yield
the following second order equation for $\psi$:
\begin{align}
&\quad\pfrac{}{x^\alpha}\left\{\left[-\ihalf e\indices{^\alpha_j}\gamma^j\psi+\frac{\rmi}{3M}e\indices{^\alpha_j}\,\sigma^{ji}\,e\indices{_i^\beta}
\left(\pfrac{\psi}{x^\beta}-\iquarter\ho\indices{^n_{m\beta}}\,\sigma\indices{_n^m}\,\psi\right)\right]\dete\right\}\nonumber\\
&=\left(\ihalf M\gamma_{k}e\indices{^k_\alpha}-\iquarter\ho\indices{^k_{l\alpha}}\sigma\indices{_k^l}\right)
\left[\ihalf e\indices{^\alpha_j}\gamma^j\psi-\frac{\rmi}{3M}e\indices{^\alpha_j}\sigma^{ji}\,e\indices{_i^\beta}
\left(\pfrac{\psi}{x^\beta}-\iquarter\ho\indices{^n_{m\beta}}\sigma\indices{_n^m}\psi\right)\right]\dete-\left(m-M\right)\psi\,\dete
\label{eq:can-dirac1d}
\end{align}
Equation~(\ref{eq:can-dirac1d}) writes in expanded form:
\begin{align*}
&\qquad e\indices{^k_\xi}\pfrac{e\indices{^\xi_k}}{x^\alpha}
\left[\ihalf e\indices{^\alpha_j}\bgamma^j\psi-\frac{\rmi}{3M}e\indices{^\alpha_j}\,\bsigma^{ji}\,e\indices{_i^\beta}
\left(\pfrac{\psi}{x^\beta}-\iquarter\ho\indices{^n_{m\beta}}\,\bsigma\indices{_n^m}\,\psi\right)\right]\dete\\
&\quad-\ihalf\pfrac{e\indices{^\alpha_j}}{x^\alpha}\bgamma^j\psi\,\dete-\ihalf e\indices{^\alpha_j}\bgamma^j\pfrac{\psi}{x^\alpha}\dete\\
&\quad+\frac{\rmi}{3M}e\indices{^\alpha_j}\,\bsigma^{ji}\,e\indices{_i^\beta}\left(
\cancel{\ppfrac{\psi}{x^\beta}{x^\alpha}}-\iquarter\pfrac{\ho\indices{^n_{m\beta}}}{x^\alpha}\,\bsigma\indices{_n^m}\,\psi
-\iquarter\ho\indices{^n_{m\beta}}\,\bsigma\indices{_n^m}\pfrac{\psi}{x^\alpha}\right)\dete\\
&\quad+\frac{\rmi}{3M}\left(\pfrac{e\indices{^\alpha_j}}{x^\alpha}\,\bsigma^{ji}\,e\indices{_i^\beta}+
e\indices{^\alpha_j}\,\bsigma^{ji}\,\pfrac{e\indices{_i^\beta}}{x^\alpha}\right)
\left(\pfrac{\psi}{x^\beta}-\iquarter\ho\indices{^n_{m\beta}}\,\bsigma\indices{_n^m}\,\psi\right)\dete\\
&=\ihalf M\bgamma_k\,e\indices{^k_\alpha}\left[\bcancel{\ihalf e\indices{^\alpha_j}\bgamma^j\psi}
-\frac{\rmi}{3M}e\indices{^\alpha_j}\,\bsigma^{ji}\,e\indices{_i^\beta}
\left(\pfrac{\psi}{x^\beta}-\iquarter\ho\indices{^n_{m\beta}}\,\bsigma\indices{_n^m}\,\psi\right)\right]\dete\\
&\quad-\iquarter\,\ho\indices{^k_{l\alpha}}\,\sigma\indices{_k^l}\left[\ihalf e\indices{^\alpha_j}\bgamma^j\psi
-\frac{\rmi}{3M}e\indices{^\alpha_j}\,\bsigma^{ji}\,e\indices{_i^\beta}
\left(\pfrac{\psi}{x^\beta}-\iquarter\ho\indices{^n_{m\beta}}\,\bsigma\indices{_n^m}\,\psi\right)\right]\dete\\
&\quad-\left(m-\bcancel{M}\right)\psi\,\dete,
\end{align*}
hence
\begin{align*}
&\quad\frac{\rmi}{3M}\Bigg[-e\indices{^\alpha_j}\,\bsigma^{ji}\,e\indices{_i^\beta}\,e\indices{^k_\xi}\pfrac{e\indices{^\xi_k}}{x^\alpha}
\left(\pfrac{\psi}{x^\beta}-\iquarter\ho\indices{^n_{m\beta}}\,\bsigma\indices{_n^m}\,\psi\right)
-\iquarter e\indices{^\alpha_j}\,\bsigma^{ji}\,e\indices{_i^\beta}\left(
\pfrac{\ho\indices{^n_{m\beta}}}{x^\alpha}\,\bsigma\indices{_n^m}\,\psi
+\ho\indices{^n_{m\beta}}\,\bsigma\indices{_n^m}\pfrac{\psi}{x^\alpha}\right)\\
&\qquad\quad+\left(\pfrac{e\indices{^\alpha_j}}{x^\alpha}\,\bsigma^{ji}\,e\indices{_i^\beta}+
e\indices{^\alpha_j}\,\bsigma^{ji}\,\pfrac{e\indices{_i^\beta}}{x^\alpha}\right)
\left(\pfrac{\psi}{x^\beta}-\iquarter\ho\indices{^n_{m\beta}}\,\bsigma\indices{_n^m}\,\psi\right)\\
&\qquad\quad+\iquarter\,\ho\indices{^k_{l\alpha}}\,\sigma\indices{_k^l}e\indices{^\alpha_j}\,\bsigma^{ji}\,e\indices{_i^\beta}
\left(\pfrac{\psi}{x^\beta}-\iquarter\ho\indices{^n_{m\beta}}\,\bsigma\indices{_n^m}\,\psi\right)\Bigg]\\
&=\rmi\,e\indices{^\alpha_j}\,\bgamma^j\left(\pfrac{\psi}{x^\alpha}-\iquarter\ho\indices{^n_{m\alpha}}\,\bsigma\indices{_n^m}\,\psi\right)-m\,\psi
+\ihalf\bgamma^j\left(\pfrac{e\indices{^\alpha_j}}{x^\alpha}
-e\indices{^\alpha_j}\,e\indices{^k_\xi}\pfrac{e\indices{^\xi_k}}{x^\alpha}\right)\psi
-\oneeighth\ho\indices{^n_{m\alpha}}\,e\indices{^\alpha_j}\left(\bgamma^j\,\bsigma\indices{_n^m}-\bsigma\indices{_n^m}\,\bgamma^j\right)\psi.
\end{align*}
The last line is converted according to:
\begin{equation*}
\oneeighth\,\ho\indices{^n_{m\alpha}}\,e\indices{^\alpha_j}\left(\bgamma^j\bsigma\indices{_n^m}-\sigma\indices{_n^m}\bgamma^j\right)
=\iquarter\,\ho\indices{^n_{m\alpha}}\,e\indices{^\alpha_j}\left(\delta^j_n\bgamma^m-\eta^{mj}\bgamma_n\right)
=\ihalf\bgamma^j\,e\indices{^\alpha_k}\,\ho\indices{^k_{j\alpha}}
\end{equation*}
which yields
\begin{align*}
&\frac{\rmi}{3M}\Bigg[-\iquarter e\indices{^\alpha_j}\,e\indices{_i^\beta}\left(\bsigma^{ji}\pfrac{\ho\indices{^n_{m\beta}}}{x^\alpha}
+\iquarter\,\ho\indices{^k_{l\alpha}}\,\sigma\indices{_k^l}\,\bsigma^{ji}\,\ho\indices{^n_{m\beta}}\right)\bsigma\indices{_n^m}\,\psi
+\iquarter e\indices{^\alpha_j}\,e\indices{_i^\beta}\,\ho\indices{^n_{m\alpha}}\left(\sigma\indices{_n^m}\bsigma^{ji}\,
-\bsigma^{ji}\,\bsigma\indices{_n^m}\right)\pfrac{\psi}{x^\beta}\\
&\qquad+\left(\pfrac{e\indices{^\alpha_j}}{x^\alpha}\,\bsigma^{ji}\,e\indices{_i^\beta}+
e\indices{^\alpha_j}\,\bsigma^{ji}\,\pfrac{e\indices{_i^\beta}}{x^\alpha}
-e\indices{^\alpha_j}\,\bsigma^{ji}\,e\indices{_i^\beta}\,e\indices{^k_\xi}\pfrac{e\indices{^\xi_k}}{x^\alpha}\right)
\left(\pfrac{\psi}{x^\beta}-\iquarter\ho\indices{^n_{m\beta}}\,\bsigma\indices{_n^m}\,\psi\right)\Bigg]\\
&=\rmi\,e\indices{^\alpha_j}\,\bgamma^j\left(\pfrac{\psi}{x^\alpha}-\iquarter\ho\indices{^n_{m\alpha}}\,\bsigma\indices{_n^m}\,\psi\right)-m\,\psi
+\ihalf\bgamma^j\left(\pfrac{e\indices{^\alpha_j}}{x^\alpha}-e\indices{^\alpha_k}\,\ho\indices{^k_{j\alpha}}
-e\indices{^\alpha_j}e\indices{^k_\xi}\pfrac{e\indices{^\xi_k}}{x^\alpha}\right)\psi.
\end{align*}
The first two lines are equivalently expressed as:
\begin{align*}
&\frac{\rmi}{3M}\Bigg[-\iquarter e\indices{^\alpha_j}\,e\indices{_i^\beta}\left(\bsigma^{ji}\pfrac{\ho\indices{^n_{m\beta}}}{x^\alpha}
+\iquarter\,\ho\indices{^k_{l\alpha}}\,\bsigma^{ji}\,\bsigma\indices{_k^l}\,\ho\indices{^n_{m\beta}}\right)\bsigma\indices{_n^m}\,\psi\\
&\qquad+\iquarter e\indices{^\alpha_j}\,e\indices{_i^\beta}\,\ho\indices{^k_{l\alpha}}\left(\sigma\indices{_k^l}\bsigma^{ji}\,
-\bsigma^{ji}\,\bsigma\indices{_k^l}\right)\left(\pfrac{\psi}{x^\beta}-\iquarter\ho\indices{^n_{m\beta}}\,\bsigma\indices{_n^m}\,\psi\right)\\
&\qquad+\left(\pfrac{e\indices{^\alpha_j}}{x^\alpha}\,\bsigma^{ji}\,e\indices{_i^\beta}+
e\indices{^\alpha_j}\,\bsigma^{ji}\,\pfrac{e\indices{_i^\beta}}{x^\alpha}
-e\indices{^\alpha_j}\,\bsigma^{ji}\,e\indices{_i^\beta}\,e\indices{^k_\xi}\pfrac{e\indices{^\xi_k}}{x^\alpha}\right)
\left(\pfrac{\psi}{x^\beta}-\iquarter\ho\indices{^n_{m\beta}}\,\bsigma\indices{_n^m}\,\psi\right)\Bigg]\\
&=\rmi\,e\indices{^\alpha_j}\,\bgamma^j\left(\pfrac{\psi}{x^\alpha}-\iquarter\ho\indices{^n_{m\alpha}}\,\bsigma\indices{_n^m}\,\psi\right)-m\,\psi
+\ihalf\bgamma^j\left(\pfrac{e\indices{^\alpha_j}}{x^\alpha}-e\indices{^\alpha_k}\,\ho\indices{^k_{j\alpha}}
-e\indices{^\alpha_j}e\indices{^k_\xi}\pfrac{e\indices{^\xi_k}}{x^\alpha}\right)\psi.
\end{align*}
The product of two $\bsigma$ matrices in the first and second line is converted according to:
\begin{equation*}
\iquarter\ho\indices{^k_{l\alpha}}\left(\sigma\indices{_k^l}\bsigma^{ji}-\bsigma^{ji}\,\bsigma\indices{_k^l}\right)
=\ho\indices{^i_{k\alpha}}\,\bsigma^{kj}-\ho\indices{^j_{k\alpha}}\,\bsigma^{ki}
\end{equation*}
which yields
\begin{align*}
&\quad\frac{\rmi}{3M}\Bigg[-\iquarter e\indices{^\alpha_j}\,\bsigma^{ji}\,e\indices{_i^\beta}\left(\pfrac{\ho\indices{^n_{m\beta}}}{x^\alpha}
+\iquarter\,\sigma\indices{_k^l}\,\ho\indices{^k_{l\alpha}}\,\ho\indices{^n_{m\beta}}\right)\bsigma\indices{_n^m}\,\psi
+e\indices{^\alpha_j}\,e\indices{_i^\beta}\left(\ho\indices{^i_{k\alpha}}\,\bsigma^{kj}
-\ho\indices{^j_{k\alpha}}\,\bsigma^{ki}\right)\left(\pfrac{\psi}{x^\beta}-\iquarter\ho\indices{^n_{m\beta}}\,\bsigma\indices{_n^m}\,\psi\right)\\
&\qquad\quad+\left(\pfrac{e\indices{^\alpha_j}}{x^\alpha}\,\bsigma^{ji}\,e\indices{_i^\beta}+
e\indices{^\alpha_j}\,\bsigma^{ji}\,\pfrac{e\indices{_i^\beta}}{x^\alpha}
-e\indices{^\alpha_j}\,\bsigma^{ji}\,e\indices{_i^\beta}\,e\indices{^k_\xi}\pfrac{e\indices{^\xi_k}}{x^\alpha}\right)
\left(\pfrac{\psi}{x^\beta}-\iquarter\ho\indices{^n_{m\beta}}\,\bsigma\indices{_n^m}\,\psi\right)\Bigg]\\
&=\rmi\,e\indices{^\alpha_j}\,\bgamma^j\left(\pfrac{\psi}{x^\alpha}-\iquarter\ho\indices{^n_{m\alpha}}\,\bsigma\indices{_n^m}\,\psi\right)-m\,\psi
+\ihalf\bgamma^j\left(\pfrac{e\indices{^\alpha_j}}{x^\alpha}-e\indices{^\alpha_k}\,\ho\indices{^k_{j\alpha}}
-e\indices{^\alpha_j}e\indices{^k_\xi}\pfrac{e\indices{^\xi_k}}{x^\alpha}\right)\psi
\end{align*}
By virtue of the identity for the product of three $\bsigma$-matrices,
\begin{equation*}
e\indices{^\alpha_j}\,\bsigma^{ji}\,e\indices{_i^\beta}\,\iquarter\,\ho\indices{^k_{l\alpha}}\,\bsigma\indices{_k^l}\,\bsigma\indices{_n^m}\,\ho\indices{^n_{m\beta}}
=e\indices{^\alpha_j}\,\bsigma^{ji}\,e\indices{_i^\beta}\,\bsigma\indices{_n^m}\,\ho\indices{^n_{k\alpha}}\,\ho\indices{^k_{m\beta}},
\end{equation*}
the generalized Dirac equation acquires the final form:
\begin{align*}
&\frac{\rmi}{3M}\Bigg[-\iquarter e\indices{^\alpha_j}\,e\indices{_i^\beta}\,\bsigma^{ji}\,\bsigma\indices{_n^m}
\left(\pfrac{\ho\indices{^n_{m\beta}}}{x^\alpha}+\ho\indices{^n_{k\alpha}}\,\ho\indices{^k_{m\beta}}\right)\psi
+e\indices{^\alpha_j}\,e\indices{_i^\beta}\left(\ho\indices{^i_{k\alpha}}\,\bsigma^{kj}-\ho\indices{^j_{k\alpha}}\,\bsigma^{ki}\right)
\left(\pfrac{\psi}{x^\beta}-\iquarter\ho\indices{^n_{m\beta}}\,\bsigma\indices{_n^m}\,\psi\right)\\
&\qquad+\left(\pfrac{e\indices{^\alpha_j}}{x^\alpha}\,\bsigma^{ji}\,e\indices{_i^\beta}+
e\indices{^\alpha_j}\,\bsigma^{ji}\,\pfrac{e\indices{_i^\beta}}{x^\alpha}
-e\indices{^\alpha_j}\,\bsigma^{ji}\,e\indices{_i^\beta}\,e\indices{^k_\xi}\pfrac{e\indices{^\xi_k}}{x^\alpha}\right)
\left(\pfrac{\psi}{x^\beta}-\iquarter\ho\indices{^n_{m\beta}}\,\bsigma\indices{_n^m}\,\psi\right)\Bigg]\\
&=\rmi\,e\indices{^\alpha_j}\,\bgamma^j\left(\pfrac{\psi}{x^\alpha}-\iquarter\ho\indices{^n_{m\alpha}}\,\bsigma\indices{_n^m}\,\psi\right)-m\,\psi
+\ihalf\bgamma^j\left(\pfrac{e\indices{^\alpha_j}}{x^\alpha}-e\indices{^\alpha_k}\,\ho\indices{^k_{j\alpha}}
-e\indices{^\alpha_j}e\indices{^k_\xi}\pfrac{e\indices{^\xi_k}}{x^\alpha}\right)\psi.
\end{align*}
Due to the skew-symmetry of $\bsigma^{ji}$, term proportional to $\bsigma\indices{_n^m}$ is actually half
the Riemann-Cartan curvature tensor in the mixed Lorentz-coordinate space representation.
(see App.~\ref{app:gamma_connection} for details).
\begin{equation*}
R\indices{^n_{m\alpha\beta}}=\pfrac{\ho\indices{^n_{m\beta}}}{x^\alpha}-\pfrac{\ho\indices{^n_{m\alpha}}}{x^\beta}
+\ho\indices{^n_{k\alpha}}\,\ho\indices{^k_{m\beta}}-\ho\indices{^n_{k\beta}}\,\ho\indices{^k_{m\alpha}}.
\end{equation*}
\subsection{Full contraction of the Riemann tensor with fundamental spinors (Dirac matrices)\label{app:riem-contr}}
From the definition of the Dirac algebra for a general contravariant metric $g^{\eta\alpha}(x)=g^{\alpha\eta}(x)$,
\begin{equation*}
\onehalf\left(\bgamma^{\eta}\bgamma^{\alpha}+\bgamma^{\alpha}\bgamma^{\eta}\right)=g^{\eta\alpha}\,\bEins,
\end{equation*}
where
$\bEins$ denotes the unit matrix in spinor space, one concludes
\begin{align*}
\bgamma^{\eta}\bgamma^{\alpha}\bgamma^{\beta}&=
\left(\bgamma^{\eta}\bgamma^{\alpha}+\bgamma^{\alpha}\bgamma^{\eta}\right)\bgamma^{\beta}-\bgamma^{\alpha}\bgamma^{\eta}\bgamma^{\beta}\\
&=2g^{\eta\alpha}\bgamma^{\beta}-\bgamma^{\alpha}\bgamma^{\eta}\bgamma^{\beta}\\
&=2g^{\eta\alpha}\bgamma^{\beta}-\bgamma^{\alpha}\left(\bgamma^{\eta}\bgamma^{\beta}+\bgamma^{\beta}\bgamma^{\eta}\right)+
\bgamma^{\alpha}\bgamma^{\beta}\bgamma^{\eta}\\
&=2g^{\eta\alpha}\bgamma^{\beta}-2g^{\beta\eta}\bgamma^{\alpha}+\left(\bgamma^{\alpha}\bgamma^{\beta}+
\bgamma^{\beta}\bgamma^{\alpha}\right)\bgamma^{\eta}-\bgamma^{\beta}\bgamma^{\alpha}\bgamma^{\eta}\\
&=2g^{\eta\alpha}\bgamma^{\beta}-2g^{\beta\eta}\bgamma^{\alpha}+2g^{\alpha\beta}\bgamma^{\eta}-\bgamma^{\beta}\bgamma^{\alpha}\bgamma^{\eta},
\end{align*}
hence
\begin{equation}\label{eq:gamma3-identity}
\bgamma^{\eta}\bgamma^{\alpha}\bgamma^{\beta}+\bgamma^{\beta}\bgamma^{\alpha}\bgamma^{\eta}
=2\left(g^{\eta\alpha}\bgamma^{\beta}-g^{\beta\eta}\bgamma^{\alpha}+g^{\alpha\beta}\bgamma^{\eta}\right).
\end{equation}
The corresponding algebra rule can be derived on the basis of~(\ref{eq:gamma3-identity}) for the product of four $\gamma$-matrices:
\begin{align}
\bgamma^{\xi}\bgamma^{\eta}\bgamma^{\alpha}\bgamma^{\beta}+\bgamma^{\beta}\bgamma^{\alpha}\bgamma^{\eta}\bgamma^{\xi}+
\bgamma^{\xi}\bgamma^{\beta}\bgamma^{\alpha}\bgamma^{\eta}+\bgamma^{\eta}\bgamma^{\alpha}\bgamma^{\beta}\bgamma^{\xi}
&=\bgamma^{\xi}\left(\bgamma^{\eta}\bgamma^{\alpha}\bgamma^{\beta}
+\bgamma^{\beta}\bgamma^{\alpha}\bgamma^{\eta}\right)+\left(\bgamma^{\beta}\bgamma^{\alpha}\bgamma^{\eta}+
\bgamma^{\eta}\bgamma^{\alpha}\bgamma^{\beta}\right)\bgamma^{\xi}\nonumber\\
&=2\bgamma^{\xi}\left(g^{\eta\alpha}\bgamma^{\beta}-g^{\beta\eta}\bgamma^{\alpha}+g^{\alpha\beta}\bgamma^{\eta}\right)
+2\left(g^{\beta\alpha}\bgamma^{\eta}-g^{\eta\beta}\bgamma^{\alpha}+g^{\alpha\eta}\bgamma^{\beta}\right)\bgamma^{\xi}\nonumber\\
&=2\left[g^{\eta\alpha}\left(\bgamma^{\xi}\bgamma^{\beta}+\bgamma^{\beta}\bgamma^{\xi}\right)
-g^{\beta\eta}\left(\bgamma^{\xi}\bgamma^{\alpha}+\bgamma^{\alpha}\bgamma^{\xi}\right)
+g^{\alpha\beta}\left(\bgamma^{\xi}\bgamma^{\eta}+\bgamma^{\eta}\bgamma^{\xi}\right)\right]\nonumber\\
&=4\left(g^{\eta\alpha}g^{\xi\beta}-g^{\beta\eta}g^{\xi\alpha}+g^{\alpha\beta}g^{\xi\eta}\right)\bEins.
\label{eq:gamma4-identity}
\end{align}
The Riemann tensor is skew-symmetric in its first and second index pair:
\begin{equation*}
R_{\xi\eta\alpha\beta}=-R_{\eta\xi\alpha\beta},\qquad R_{\xi\eta\alpha\beta}=-R_{\xi\eta\beta\alpha}.
\end{equation*}
Thus
\begin{align}
R_{\xi\eta\alpha\beta}\left(\bgamma^{\xi}\bgamma^{\eta}\bgamma^{\alpha}\bgamma^{\beta}+\bgamma^{\beta}\bgamma^{\alpha}\bgamma^{\eta}\bgamma^{\xi}+
\bgamma^{\xi}\bgamma^{\beta}\bgamma^{\alpha}\bgamma^{\eta}+\bgamma^{\eta}\bgamma^{\alpha}\bgamma^{\beta}\bgamma^{\xi}\right)
&=4R_{\xi\eta\alpha\beta}\left(g^{\eta\alpha}g^{\xi\beta}-g^{\beta\eta}g^{\xi\alpha}+\cancel{g^{\alpha\beta}g^{\xi\eta}}\,\right)\bEins\nonumber\\
&=-4\left(R_{\xi\eta\beta\alpha}\,g^{\eta\alpha}g^{\xi\beta}+R_{\xi\eta\alpha\beta}\,g^{\beta\eta}g^{\xi\alpha}\right)\bEins\nonumber\\
&=-4\left(R\indices{^\beta_{\eta\beta\alpha}}\,g^{\eta\alpha}+R\indices{^\alpha_{\eta\alpha\beta}}\,g^{\beta\eta}\right)\bEins\nonumber\\
&=-4\left(R\indices{_{\eta\alpha}}\,g^{\eta\alpha}+R\indices{_{\eta\beta}}\,g^{\beta\eta}\right)\bEins=-8R\indices{_\eta^\eta}\bEins\nonumber\\
&=-8R\,\bEins.
\label{eq:gamma4a-identity}
\end{align}
For zero torsion, the Riemann tensor has the additional symmetries:
\begin{equation*}
R_{\xi\eta\alpha\beta}=R_{\alpha\beta\xi\eta}=R_{\beta\alpha\eta\xi},\qquad R\indices{^\alpha_{\eta\alpha\beta}}=R_{\eta\beta}=R_{\beta\eta},\qquad
R_{\xi\eta\alpha\beta}+R_{\xi\alpha\beta\eta}+R_{\xi\beta\eta\alpha}=0.
\end{equation*}
By virtue of these symmetries, the left-hand side of Eq.~(\ref{eq:gamma4a-identity}) simplifies to:
\begin{align*}
&\quad\,R_{\xi\eta\alpha\beta}\left(\bgamma^{\xi}\bgamma^{\eta}\bgamma^{\alpha}\bgamma^{\beta}+\bgamma^{\beta}\bgamma^{\alpha}\bgamma^{\eta}\bgamma^{\xi}+
\bgamma^{\xi}\bgamma^{\beta}\bgamma^{\alpha}\bgamma^{\eta}+\bgamma^{\eta}\bgamma^{\alpha}\bgamma^{\beta}\bgamma^{\xi}\right)\\
&=2R_{\xi\eta\alpha\beta}\left(\bgamma^{\xi}\bgamma^{\eta}\bgamma^{\alpha}\bgamma^{\beta}+
\bgamma^{\xi}\bgamma^{\beta}\bgamma^{\alpha}\bgamma^{\eta}\right)\\
&=2R_{\xi\eta\alpha\beta}\left(\bgamma^{\xi}\bgamma^{\eta}\bgamma^{\alpha}\bgamma^{\beta}
-\bgamma^{\xi}\bgamma^{\eta}\bgamma^{\beta}\bgamma^{\alpha}-\bgamma^{\xi}\bgamma^{\alpha}\bgamma^{\eta}\bgamma^{\beta}\right)\\
&=2R_{\xi\eta\alpha\beta}\left[2\bgamma^{\xi}\bgamma^{\eta}\bgamma^{\alpha}\bgamma^{\beta}
-\bgamma^{\xi}\left(2g^{\alpha\eta}\,\bEins-\bgamma^{\eta}\bgamma^{\alpha}\right)\bgamma^{\beta}\right]\\
&=2R_{\xi\eta\alpha\beta}\left(3\bgamma^{\xi}\bgamma^{\eta}\bgamma^{\alpha}\bgamma^{\beta}
-2g^{\alpha\eta}\bgamma^{\xi}\bgamma^{\beta}\right)\\
&=6R_{\xi\eta\alpha\beta}\,\bgamma^{\xi}\bgamma^{\eta}\bgamma^{\alpha}\bgamma^{\beta}
+4R_{\eta\xi\alpha\beta}\,g^{\alpha\eta}\bgamma^{\xi}\bgamma^{\beta}\\
&=6R_{\xi\eta\alpha\beta}\,\bgamma^{\xi}\bgamma^{\eta}\bgamma^{\alpha}\bgamma^{\beta}
+2R_{\xi\beta}\left(\bgamma^{\xi}\bgamma^{\beta}+\bgamma^{\beta}\bgamma^{\xi}\right)\\
&=6R_{\xi\eta\alpha\beta}\,\bgamma^{\xi}\bgamma^{\eta}\bgamma^{\alpha}\bgamma^{\beta}
+4R_{\xi\beta}\,g^{\xi\beta}\,\bEins\\
&=6R_{\xi\eta\alpha\beta}\,\bgamma^{\xi}\bgamma^{\eta}\bgamma^{\alpha}\bgamma^{\beta}+4R\,\bEins\\
&\stackrel{\text{(\ref{eq:gamma4a-identity})}}{=}-8R\,\bEins.
\end{align*}
The full contraction of the Riemann tensor with Dirac matrices is thus obtained for zero torsion as:
\begin{equation}\label{eq:Riemann-identity}
-R_{\xi\eta\alpha\beta}\,\bsigma^{\xi\eta}\,\bsigma^{\alpha\beta}=
R_{\xi\eta\alpha\beta}\,\bgamma^{\xi}\bgamma^{\eta}\bgamma^{\alpha}\bgamma^{\beta}=-2R\,\bEins
\quad\Leftrightarrow\quad R_{\xi\eta\alpha\beta}\,\gamma\indices{^a_c^\xi}\,
\gamma\indices{^c_d^\eta}\,\gamma\indices{^d_e^\alpha}\,\gamma\indices{^e_b^\beta}=
-2R_{\xi\eta\alpha\beta}\,g^{\xi\alpha}\,g^{\eta\beta}\,\delta_b^a,
\end{equation}
with spin indices denoted by Latin letters.
Equation~(\ref{eq:Riemann-identity}) is consistent with the trace identity of four Dirac matrices:
\allowdisplaybreaks[0]
\begin{align*}
R_{\xi\eta\alpha\beta}\,\mathrm{Tr}\left\{\bgamma^{\xi}\bgamma^{\eta}\bgamma^{\alpha}\bgamma^{\beta}\right\}
&=4R_{\xi\eta\alpha\beta}\left(\,\cancel{g^{\xi\eta}g^{\alpha\beta}}-g^{\xi\alpha}g^{\eta\beta}+g^{\xi\beta}g^{\eta\alpha}\right)\\
&=-8R\indices{^\alpha_{\eta\alpha\beta}}\,g^{\eta\beta}\\
&=-8R\stackrel{\text{(\ref{eq:Riemann-identity})}}{=}-2R\,\,\mathrm{Tr}\left\{\bEins\right\}\quad\text{as}\quad\mathrm{Tr}\left\{\bEins\right\}=4.
\end{align*}
Equation~(\ref{eq:Riemann-identity}) further simplifies Eqs.~(\ref{eq:gen-dirac-mc}) resp.~(\ref{eq:gen-dirac-mc-2}) according to:
\begin{equation*}
\frac{1}{24M}\sigma^{\alpha\beta}\sigma^{nm}\,R_{nm\alpha\beta}
=-\frac{1}{24M}R_{\xi\eta\alpha\beta}\,\bgamma^{\xi}\bgamma^{\eta}\bgamma^{\alpha}\bgamma^{\beta}=\frac{1}{12M}\,R\,\bEins,
\end{equation*}
and thus provides the additional scalar mass-like term in the generalized Dirac equation~(\ref{eq:dirac-final-notorsion}).
\end{document}